\documentclass[12pt,preprint]{aastex}

\begin{document}

\title{Chandra Observations of Associates of $\eta$ Car:  I. Luminosities
  \altaffilmark{1} }

\author{Nancy Remage Evans, Frederick D. Seward, Miriam I. Krauss, }

\author{Takashi Isobe, Joy Nichols, Eric M. Schlegel, and Scott J. Wolk }

\affil{Smithsonian Astrophysical Observatory, MS 4,
 60 Garden St., Cambridge, MA 02138}


\altaffiltext{1}{Based on observations made with the Chandra
X-Ray Observatory.}







\begin{abstract} 

The region around the $\eta$ Car nebula has three OB associations, 
which contain a Wolf-Rayet star and several massive O3 stars.             
An early Chandra ACIS-I image was centered on $\eta$ Car    
and includes Trumpler 16 and  part of Trumpler 14. The Chandra image         
confirms the well-known result that O and very early B stars          
are X-ray sources with 
L$_X$ $\simeq$ 10$^{-7} \times$ L$_{bol}$ over an X-ray 
luminosity range of about 100.  Two new anomalously strong X-ray 
sources have been found among the hot star population,         
Tr 16-244, a heavily-reddened O3 I star, and Tr 16-22, a         
heavily-reddened O8.5 V star. Two stars have an unusually large
L$_X$/L$_{bol}$: HD 93162, a Wolf-Rayet star (and possible binary), 
and Tr 16-22, a possible colliding wind binary.  
In addition, a population of sources associated with cool stars is 
detected. In the color-magnitude diagram, 
 these X-ray sources sit above the sequence of field stars in the 
Carina arm. The OB stars are on average   more X-ray luminous than the 
cool star  X-ray sources. X-ray sources among A stars
have similar X-ray luminosities 
to cooler stars, and may be due to cooler
companions. Upper limits are presented for B stars which are
not detected in X-rays.  These upper limits  are also the 
upper limits for any cool companions which the hot stars
may have.   Hardness 
ratios are presented for the most luminous sources in  
bands 0.5 to 0.9 keV, 0.9 to 1.5 keV, and 1.5 to 2.04 kev. 
The available information on the binary nature of the  hot stars is discussed, 
but binarity does not correlate  with  X-ray strength in a simple way.

\end{abstract}


\section{Introduction}

The challenge of X-rays in O stars was posed by Einstein 
X-ray Observatory observations 
of the Carina nebula (Seward et al. 1979; Seward and Chlebowski, 1982)
and Cyg OB2 (Harnden et al. 1979).
Four particularly strong sources were found in Tr14 and Tr 16: $\eta$
Car itself, HD 93129A,B (O3 If), HD 93162 = WR25, a Wolf-Rayet 
star, and HD 93250 (O3V).  All the bright O stars in the Einstein field were 
detected in X-rays.  Among the clues to the source 
of X-rays was  that L$_X$ is approximately seven orders of 
magnitude smaller than L$_{bol}$ for all O stars.
  The W-R star WR25 is unusually
strong in X-rays compared with two other W-R stars  in the 
region.  

Several X-ray studies have been made 
subsequently, 
including ROSAT images (Corcoran et al. 1995) and
Crowther et al. (1995a), both of whom 
confirmed that WR25 is unusually bright 
for a W-R star.

The region around $\eta$ Car is of considerable interest.  
$\eta$ Car itself is arguably the most massive star in the 
galaxy, with a history of luminosity variations. A Hubble Space 
Telescope image revealed a bipolar structure in the center of 
the nebula.
The Carina nebula contains the most massive stars in the 
Galaxy with initial masses up to 120 M$_\sun$.
Walborn (1971) invented the spectral classification ``O3" 
because several stars in this region 
are hotter than existing MK classifications.
Recent studies include the 
 Massey and Johnson study (1993, MJ93 hereafter) of reddening
from spectroscopy and photometry, and the proper motion study of
Cudworth et al. (1993), which provides membership probability. 
The clusters Tr14 and Tr16 provide a snapshot of evolution of the 
most massive stars known.  The evolutionary sequence in order of 
decreasing mass goes from $\eta$ Car through the O3 If star through
O3 V stars.  A particularly perplexing question in this scenario 
is how the W-R star WR25 fits in.  This is discussed at some length
by MJ93.  It should be more evolved than the O3 stars;
however, it is less luminous.  This statement, of course, depends 
on the bolometric corrections, which are difficult to determine for 
 O stars and especially difficult for W-R stars.

WR25 is classified as WN7+abs by MJ93.  WN7+abs stars
belong to the nitrogen sequence of W-R stars which are the most luminous and
least chemically evolved as mass loss strips off outer layers of the original
star. They still retain some hydrogen, and the ``abs" designation indicates
that there are absorption components in the upper Balmer lines. 
WN7+abs stars differ from Of stars only in having a stronger wind. 
As noted by Crowther et al. (1995b), WN7+abs stars 
are found only in the youngest clusters, in 
contrast to WN7 stars. The scenario put forward 
by Crowther et al. (1995 a and b), based on their spectroscopy, envisions 
that WN7+abs stars are part of the most massive sequence of stars,
immediately following the evolutionary stage of Of stars, but preceding WN7
stars.  (WN8
stars belong to a less massive sequence containing stars
which have already been luminous blue
variables or red supergiants.)  
This scenario accords with the selection of
objects in Tr16 and Tr14 (WN7+abs, O3 If*, O3 V, and cooler stars).

 Tr 16 and Tr 14 contain some of the youngest and most massive 
stars known, and  have been well-studied.  For the same reason,
 many of their parameters are extreme, and also
controversial.  
 The distance, for instance, adopted by Davidson 
and Humphreys (1997) to $\eta$ Car itself is 2.3 kpc.  
MJ93 find 3.2 kpc from extensive spectroscopy and 
photometry.  Among the problems in determining the distance 
to Tr 14 and Tr 16 are the form of the reddening law, and 
whether the two clusters have the same age and distance.  Walborn 
(1995) illustrates quantitatively how the reddening law affects 
the derived distance.  For R = A$_V$/E(B-V) = 3.5, his distance
to Tr 16 is 2.5 kpc. Crowther et al. (1995b)  use 2.6 kpc in their 
discussion of the Wolf-Rayet in Tr14, star WR25.  
We adopt 2.5 kpc for this study.

Given the dramatic appearance of the nebulosity in optical images
of the cluster, the reddening within Tr14 and Tr16 is surprisingly
uniform.  The values of E(B-V) for the hot stars from MJ93
 have a scatter of approximately $\pm$ 0.2 mag, centered
approximately on 0.55 mag.  The form of the reddening law in the 
region, however, has been debated.  This is 
discussed further in section 6.    
Bolometric corrections are similarly problematic in comparing 
very luminous hot stars with W-R stars.      

A good overview of X-ray studies of O stars and Wolf-Rayet 
stars in the pre-Chandra era is provided by Willis and Crowther (1996).
The search for correlations between X-ray luminosity and other 
parameters has been elusive, and is summarized by Chlebowski (1989).

  The observations of $\eta$ Car itself 
from the early Chandra image have already been 
discussed by Seward et al. (2001).  We now turn to the stars in the 
surrounding region.  The discussion will be broken into two papers.
The current paper discusses the source counts, the luminosity distributions,
and hardness ratios.  The second paper will discuss the low-resolution spectra
of the strongest sources.  The current paper is broken up into 
sections on the observation, detected sources, cool stars, luminosity 
distributions, X-ray to bolometric luminosity ratios, diffuse emission,
upper limits, and hardness ratios.

\section{Observations}

The early Chandra science image placed $\eta$ Car at the focus 
of the I3 chip of the of the ACIS-I detector.
  Instrumental details about the
Chandra satellite are given in Weisskopf et al. (2002).

Unfortunately, the $\eta$ Car 
observation was obtained during the period 
when the energy resolution of the detector was degrading.
It was also obtained at an instrumental temperature which has not been 
fully calibrated.  This means that 
 the ACIS instrumental low-resolution spectra have 
uncertainties in the energy scale.  The intrinsic interest of the 
sources in this image, however, means that there is insight to 
be gained from an analysis of the image despite these limitations. 

Fig. 1 shows the image on the ACIS-I chips. 
$\eta$ Car itself is near the center of the image, with stars in Trumpler
16 particularly prominent to the south.  Trumpler 14 is in the 
extreme upper right.  The identification  
 of several of the sources (hot stars) is provided in Fig. 1.

The data were analysed
using the CXC CIAO (Chandra Interactive 
Analysis of Observations)  software.  
The data were filtered on standard ASCA grades and checked for times
of poor aspect (pointing stability); 
none were found.  There were no significant background
flares during the observation, giving a total good exposure time of
9.63 ksec.

Sources were detected using
the {\tt WAVDETECT} program which performs a Mexican hat wavelet 
decomposition and reconstruction.\footnote{See The 
{\it CIAO} Detect Manual, Chapter 13 
http://cxc.harvard.edu/ciao/download/doc/detect\_html\_manual/ .}
{\tt WAVDETECT} was run using scales of 
size 1, 2, 4, 8, and 16 pixels on an image
which had been energy filtered to contain data from 300 to 6000 eV.
These choices maximized the number of total detections returned while
minimizing the number of spurious detections.

The data were then split into three energy bands: 0.5-0.9, 0.9-1.5,
and 1.5-2.04 keV.  An exposure map was computed for the observation 
to take into account vignetting and chip gaps. A mean energy of 
1.2 keV was used in generating  the exposure map, which was 
 normalized to its aimpoint value.
Source counts were extracted using the regions supplied 
by  {\tt WAVDETECT}.  
  These counts were divided by the normalized
mean exposure value, and background-subtracted using data from an
annulus centered on the source region.  
 Sources within the background region were 
subtracted prior to this computation and the area was adjusted accordingly.



The list of detected sources was culled in the following way to 
produce a list of significant sources.
  The distance off-axis for 
each source was calculated and for this distance the size of the 
point spread function (PSF) was determined.  See, for instance,
Fig. 6.4 in the Chandra
Proposers' Observatory Guide (2000). (Specifically, 
  a fit by J. Grimes
[2001, private communication] for 80\% encircled energy at 1.5 keV
was used.)
Background counts were measured in two areas of the image and 
found to be 0.045 counts per square arcsec.  The background within 
the psf was calculated for each radius.  For this background, 
the number of counts required for a 3 $\sigma$ detection 
was computed using the approximate 
formula from Gehrels (1986, p. 339, ``recommended approximation").   
Using this approach, for 
instance, an on-axis source requires approximately 6 counts  to 
be significant; a source 7 arcmin off-axis requires approximately 
12 counts.  

The resulting sources are listed in Tables 1 to 4. separated by
their optical identifications  into OB stars (Table 4 of MJ93)
and other sources.
The other sources in  Table 2 lists  
are predominately pre-main sequence stars.  
Sources within the $\eta$ Car nebula have been omitted from the tables.
 As well as 
total counts from 0.5 to 2.04 keV, counts were measured in 
3 energy bands, 0.5 to 0.9 keV, 0.9 to 1.5 keV, and 1.5 to 2.04 keV, 
referred to as soft (S), medium (M), and hard (H) bands. 
  Errors are computed by {\tt DMEXTRACT} using Poisson 
statistics.  
Tables 1 and 2 also list the radial distance R from the aimpoint 
(in arcmin).  

Tables 3 and 4 list quantities derived from source counts, again divided into 
OB stars and other sources.  Hardness ratios have been calculated from the 3 
energy bands:

$$ HR_{MS} = {(M-S)\over{M+S}} $$ 

$$ HR_{HM} = {(H-M)\over{H+M}} $$

Errors have been propagated from the errors on the energy bands.

   In addition, fluxes and luminosities have been calculated from 
total count rates.  PIMMS \footnote{Portable Interactive Multi-Mission 
Simulator, http://cxc.harvard.edu/toolkit/pimms.jsp }
 was used to calculate a conversion from 
counts to energy.  The count rates were calculated on the basis of 
9.630 ksec of good time.   A Raymond-Smith plasma model spectrum 
with a log temperature of  6.65 was used, a spectrum 
typical of a stellar source found in young associations.  
The extinction is 
fairly uniform over the field.  A column of 
N$_H$ of 3 x 10$^{21}$ cm$^{-2}$ 
was adopted, corresponding to E(B--V) = 0.52 mag, which is typical
of the field (MJ93).  With these parameters, a conversion factor 
of 1.4 x 10$^{-11}$ ergs cm$^{-2}$ sec$^{-1}$ per count per sec
was derived.  Conversion factors differed by less than a factor of 
two for a   range of log T from 6.4 to 7.1. 
 Luminosities were derived  from fluxes 
using a distance of 2.5 kpc.

\subsection{Hot Stars} 

Figure 2 shows the ACIS-I image of $\eta$ Car overlaid with 
the  detected sources.  
Extensive optical work has been done on the clusters in the vicinity of
$\eta$ Car, most recently by MJ93 and Cudworth,
Martin, and DeGioia--Eastwood (1993).  Table 4 in MJ93, 
 the brightest and bluest stars near $\eta$ Car, 
 has been used to identify O and B stars with X-ray 
sources (shown as circles).  Sources which are not identified with 
O and B stars are shown as diamonds.  (Note that the increase in the PSF 
 of sources which are farther off axis is not 
reflected in the plotting symbols.)  Presumably most of the 
X-ray sources which are not O stars are pre-main sequence stars, 
although very rarely they could be background objects, or even extragalactic
sources.  

Figure 3 shows the  ACIS-I image of $\eta$ Car with the O and early 
B stars from the MJ93 list marked.  It is clear
many of the X-ray sources are early type stars, but many are not.
Furthermore, many of the stars on the O and early B star list are 
not X-ray sources.   

Figure 4 shows a plot of V magnitude vs spectral type  of
the O and early B stars from the list of MJ93. 
  (A few of the hot stars are not plotted because 
spectral types are not available.) The observed V magnitude has 
not been corrected for reddening. However,  extinction is fairly 
uniform over the field, so the pattern is not changed much if 
V$_0$ is used instead.  An exception to this statement is the 
O3 star at 10.78 mag.  This star was recognized by MJ93 
to be unusually heavily reddened (MJ 257), 
and will be discussed further
below.  The  06.5 star which is shown as  not detected 
lies in the extreme upper right corner of the ACIS-I image
(Fig. 1 or 2), considerably off-axis, where a modest 
source could easily fail our detection criterion.  With this 
exception, the dividing line between detections and nondetections 
is very sharp at just approximately B0.  This confirms previous 
results for O stars.

There is one B1.5 star which is detected in X-rays.  A sole exception 
like this could easily result from a later companion.  It will be 
discussed in section 10.3.

Figure 5 shows the total X-ray counts (Table 1) as a function of 
spectral type for the hot stars.

\subsection{Cool Stars}

The field corresponding to the Chandra image of $\eta$ Car
was observed January 9-10, 2001 using the 0.9 meter telescope
at CTIO.  The Tek-2K CCD has a 13.5\arcmin\ field of view.  First, four
pointings, arrayed in a cruciform pattern with $\eta$ Carina near the
edge of each field,  were used to observe the
entire field. Johnson B and V filters and a Kron-Cousins  I 
filter were used, with exposures of
2 minutes each. This gives a limiting magnitude of about 19 at V.
We performed an additional exposure of 10 minutes with each filter, 
centered on $\eta$ Carina.  This gave us a limiting magnitude of about
20.5 at V within 6.75\arcmin\ of the aimpoint.  

Basic data reduction was handled using the {\tt quadproc} package within IRAF
 \footnote{IRAF is distributed by                                               
  the National  Astronomical Observatories, which is operated by                 
   the Association of Universities for Research in Astronomy, Inc.,               
    under contract to the National Science Foundation.} 
to bias correct and flat-field the data. 
 Fluxes for all standards and  target stars                                                                    
 were determined using the {\tt fitpsf} tool within the IRAF 
 {\tt apphot} package. Using     
 this tool fluxes were derived from the analytic fit to each 
 individual psf in    
 lieu of simple aperture photometry.                                             
 Stellar magnitudes and colors were put on the standard system using             
 the {\tt photcal} package.                         
 There were three
difficulties encountered during data analysis. First, the region has a highly
variable diffuse background. Second, $\eta$ Carina and other bright stars make
photometric measurements of stars in their proximity difficult. Finally the
region is quite crowded, and often  faint stars have overlapping PSFs.
Optical data are presented in Table 5, ordered by source number from 
Tables 2 and 4.  Problems  are noted, as appropriate according 
to the code at the end of the table. Photometric errors were
determined using fits to the standard stars and by comparing observed values
of target stars observed multiple times in separate pointings. This gives a
more realistic assessment of the errors given the highly variable and confused
nature of the field. Errors are less than 5\% at V and less than 8.3 \% in the
colors down to the completeness limits.  These  errors are somewhat 
conservative because a star                                                                            
near the center of one field is on the edge of the comparison field, thus the   
PSFs may differ greatly. 

The central image with the X-ray sources overlaid is shown in Figure 6.
The resulting color magnitude diagram is shown in Figure 7. In this figure,
X--ray sources near Trumpler 16 are shown as asterisks, those near Trumpler 14
are shown as open circles. Other sources are shown as small dots. There are
6219 stars plotted. All stars detected in B, V, and I and brighter than our
completeness limit in the deep center field exposure are plotted. Fig. 7 shows
no significant difference between the age inferred for the Trumpler 16 sources
compared with those near Trumpler 14, although the statistics are small. The
figure demonstrates that the color--magnitude diagram easily distinguishes
stars of Carina arm of the galaxy from nearby stars. The X-ray sources are all
in a band about 1 magnitude wide, located about one magnitude above the
field stars
in the Carina arm. The well defined sequence of field stars (dots) results
because they, like the cluster stars, are concentrated in the Carina arm.
Heavy obscuration behind the clusters restricts background contamination. The
large spread among the stars is similar to that seen in Orion and is common
among stars less than 1 Myr old when small differences in age can lead to
large differences in luminosity. About 100 stars in this region of the 
HR diagram are not detected in X-rays.
 There are a few X-ray sources about 0.75 magnitudes above
the upper envelope of the X--ray population; we interpret these as binary
stars. The reddest stars have V-I of about 2. Given A$_V$ = 1.95, the
dereddened V-I at the V=19 is 1.28. At an age of 1 Myr, this corresponds to a
mass of about 1.15 M$\sun$ (Baraffe et al. 1998).

\section{Luminosity Distribution}  

Figure 8 shows the X-ray luminosity 
distribution of the O and early B
stars (Table 3).  Fig. 9 shows the luminosity distribution for all 
sources which are not identified as OB stars (Table 4),
the pre-main sequence stars.  A or B 
stars with later companions might be included, as could any extra-galactic
objects, but Fig. 9 is heavily dominated by late type
stars.  
The luminosity distribution is distinctly different between these 
two groups.  Most of the OB stars have luminosities greater than 
10$^{31}$ ergs s$^{-1}$.  The cooler 
stars have luminosities strongly
peaked at 10$^{31}$ ergs s$^{-1}$, with
 the luminosity distribution of cool stars (Fig. 9)   
 truncated at the faint end because of the relatively shallow 
X-ray image.  As discussed below in Section 8, the depth of 
the X-ray exposure allowed all O stars near the image center to be
detected.

\section{X-Ray Luminosity Fraction}

\subsection{Hot Stars}

In this section we explore the efficiency of X-ray production from hot stars
(Table 6).  
  The first column is the optical identification,
taken from Table 4 of MJ93.  The second column 
is the X-ray source number from Table 3. Columns 3 and 5
are the V and E(B-V) (in magnitudes), again from MJ93.
Column 4 is the spectral type.  The reader is referred to 
Table 4 in MJ93 
for the comparison of spectral types from different 
sources and in some cases more complex spectral types.
Recently Walborn et al. (2002) have comprehensively
 reexamined the spectral classification of  
 luminous stars.   They
 make small adjustments to some classifications, including 
 introducing new classes O2 and O3.5.  The new classifications 
 for stars in Table 5 are HD93129A: O2 If*, HD93129B: O3.5 V((f+))
 HD93250: O3.5 V((f+)),  HD93205: O3.5 V((f+)),  and
  HDE303308: O4 V((f+)).
 This finer gradation  would only mildly affect the figures below.    
The sixth column is the bolometric magnitude,  taken from 
Table 3 in Massey et al. (2001) where available, made fainter by
0.56 mag because of the different distance assumed.   
 (The X-ray to bolometric luminosity ratio which we use below, of  course, 
does not depend on the distance.)  For other stars, bolometric
corrections were computed from the same formulae (Massey et al.
2000). The final column is the log of the ratio of X-ray luminosity
to the bolometric luminosity.  Some of the sources in Table 6 have 
been identified in Fig. 1.

While the reddening through the cluster field is reasonably 
uniform (MJ93), there been studies that have 
suggested that the reddening law is unusual (e.g. Herbst, 1976).
Other studies have refuted this, or concluded that only some 
objects have unusual reddening laws (Turner and Moffatt, 1980;
Tapia et al. 1988).  

One case where the reddening law does appear to be unusual is the 
W-R star HD 93162 = WR25, although the unusual spectrum 
of the W-R star presents a problem in reddening determination.  
Crowther et al. (1995b) discuss the flux distribution from IUE 
through the IR and conclude that the reddening law for WR25 is 
unusual [E(B-V) = 0.66 with R = 4.6].  This is the reddening that 
we adopt.  The unusual spectrum also complicates the determination
of the bolometric correction.  We adopt the bolometric magnitude
from Crowther et al. (1995a), -10.1 mag.  (A small correction has 
been included for the difference between the distance they
use and the one used here.)

There are a few unusual stars among the hot X-ray sources.  
Tr 16-244 is a faint star 
 with an early spectral type 
(O3-O4 If).  MJ93 took particular care in their 
observations because of this, and because the star lies  close to the 
bright W-R star HD 93162.  
  Tr 16-244 is marked in Fig. 1, and the 
W-R star is just to the right.  The reddening implied for Tr 16-244 by 
the spectral type and photometry of MJ93
is larger than other stars in the cluster, with  E(B-V) = 
0.94 mag (using the relations in Massey et al. 2000).  Tr16-244 
and HD 93162 both lie very close to the band of obscuration 
in the visual image.  However, the W-R star is between Tr16-244
and the dust lane. 
 While reddening is difficult to determine 
for W-R stars, the value found by Crowther et al. (1995b)
for the total extinction in the visual   of the Wolf-Rayet
star, (A$_V$ = 3.0 mag) is very close to that for Tr16-244.
  Thus, it is not clear whether the large reddening for those two 
  stars is  caused by the dust lane, or could have a circumstellar
  component (for each star).
 For Tr 16-244, since the spectral type is the same as for 
HD 93129AB, the same temperature and bolometric correction were used 
in creating Table 6. The resulting M$_{bol}$ is -8.6 mag, although 
there are obviously several sources of uncertainty in this number.

The other hot star with surprising photometry 
 is Tr16-22.  The V magnitude is  faint for a 
spectral type of O8.5 V, suggesting a high reddening. 
Tr16-22  is very close to the dust lane that cuts diagonally
 through the lower left of the image (Fig. 1), so  the 
 foreground absorption may be particularly large for this star.  
 The spectrum 
MJ93 show is similar to other O8.5 V spectra, except that the 
lines are slightly shallower and broader. However, the 
value of Q from the photometry gives rather unconvincing values for 
temperature and bolometric correction.  Comparing the B-V and U-B 
values from MJ93 with those of Herbst (1976) and 
Feinstein (1969) shows that
the agreement between the latter two is much better. 
  For this reason, we adopt 
 E(B-V) = 0.78 mag from the mean B-V and Q of Herbst and Feinstein. 
 This provides a bolometric magnitude very close to the two
 O8.5 V stars Tr16-115 and Tr16-3, so we adopt the bolometric 
 magnitude from these stars (-6.8 mag).  

Fig. 10 shows the L$_X$/L$_{bol}$ ratio for the stars in Table 6
as a function of spectral type.
 An approximate adjustment to  the X-ray 
luminosities of HD 93162, Tr 16-244, and Tr 16-22 of a factor 
2.45 was made because of their larger reddening
which would correspond to an E(B-V) = 0.90 mag
(estimated using PIMMS).
  
As is known from previous studies, for most stars in the O and early B
spectral range,  the L$_X$/L$_{bol}$ ratio
remains constant at -7 .  There 
are 3 exceptions to this in Fig. 10.  The W-R star HD 93162 has been known
since the X-ray discovery paper (Seward et al. 1979)
to be unusually strong in X-rays
as compared with other W-R stars.  This has been particularly 
surprising as it has been thought to be a single star.  Binaries
with components this massive are often colliding wind binaries
and hence stronger X-ray sources than single stars.  However, 
the recent W-R catalogue by van der Hucht (2001) lists for the first 
time a possible binary indicator, ``diluted emission lines". 
A companion bright enough to be detected photometrically would
probably be luminous enough to have a wind which would enhance 
X-ray production.  

Tr16-22 (source 55) also has an unusually
large X-ray luminosity in comparison with its bolometric luminosity
(L$_X$/L$_{bol} \simeq$ -5.8).
As noted above, it has a larger than average absorption, possibly
because it is on the edge of the dust lane.  However, 
the  large X-ray luminosity suggests the source
itself may have unusual characteristics beyond simple foreground
reddening.  The X-ray luminosity of the source 
 (L$_X$ = 7.9   $\pm$ 0.5  x 10$^{31}$ erg sec$^{-1}$ in Table 3 before the 
adjustment for high reddening) makes it one of
the brightest O stars observed, approximately a factor of 10 larger 
than other O8.5 V stars.  

The third source with a large L$_X$/L$_{bol}$ ratio is Tr16-11, 
the B1.5 V star (source 141 in Table 1).  Fig 10 shows that 
a B star would not even be detected unless it is unusually strong
in X-rays. The luminosity 
(L$_X$ =  1.4 $\pm$ 0.5  x 10$^{31}$  erg sec$^{-1}$) is  within the 
range of the cool star distribution (Fig. 9) so a late companion
is a possibility.  MJ93 display its optical spectrum, 
which looks very like others of similar spectral type, except that 
perhaps the hydrogen lines are slightly broader. 

The spread in L$_X$/L$_{bol}$ ratio in Fig. 9 is particularly 
large among the O3 stars.  The two with the largest ratio are
HD 93250 and Tr 16-244.  

\subsection{Cool Stars} 

The X-ray production in cool stars was similarly investigated using the X-ray
luminosity (Table 4) and the optical luminosity from the data in Table 5. In
this case, the typical extinction for the clusters [E(B--V) = 0.52 mag] was
assumed for all the cool stars, as was the standard distance. Bolometric
corrections were interpolated Table 15.3.1 in {\it Astrophysical Quantities}
(Cox, 1998).  The V-I (Johnson) in that table was transformed to 
the Kron-Cousins system using the equations provided by Bessell (1979).  
The resulting plot of log L$_X$ and 
log (L$_X$/L$_{bol}$) as functions of V--I are 
are shown in Figs. 11 and 12.  For comparison,
the colors of A0V, F0 V, and K7 V stars (from Bessell, 1979) are 
shown with the reddening of the clusters.  Note that in this 
relatively short exposure we only detect a fraction of the 
stars in the cluster. For instance presumably many M stars are
not detected.

The most informative comparison is with comparable figures for 
the Orion Nebula Cluster (ONC, Chandra results in Feigelson, 
et al., 2002; ROSAT results in Gagne, et al. 1995). Stars in 
the two regions are similar in age, although those in Tr 14--Tr 16 
are slightly younger (Meynet, et al., 1993; Massey, et al., 2001).
Tr14--Tr16 contain more massive stars than the ONC.
 Some X-ray sources among A stars are detected in both clusters.  
Feigelson, et al. argue that because the L$_X$ distribution in 
late B and A stars is the same as for cooler stars, X-ray sources
among late B and A stars can be explained by less massive 
companions.  Fig. 11 shows the same result for Tr 14--Tr 16.

\section{Diffuse Emission}

To illustrate diffuse emission 
in the field, a smoothed 0.3-7 keV image was created 
using the CIAO adaptive smoothing tool {\tt CSMOOTH}. 
The result is shown in Fig. 13, which
 indicates broad, diffuse emission from much of the
nebula. We investigated two circular regions 2.5$^{\prime}$ in
diameter marked in the figure; 
one  centered 4.0$^{\prime}$ S of $\eta$ Car and one 
centered 7.5$^{\prime}$ S of HD93129 A/B.  The figure shows 7
unresolved sources in the first region, accounting for 30\% of the
emission after background subtraction.  There are no unresolved
sources in the 2nd region.

After subtracting  bright stars, the residual above background
in each region is $\approx$ 750 counts, or $\approx$ 150 counts 
arcmin$^{-2}$.  This corresponds to a surface brightness of $\approx
2.5 \times 10^{-13}$ ergs cm$^{-2}$ s$^{-1}$ arcmin$^{-2}$ (unabsorbed).  
The luminosity of this region, if spherical, 
is 3 $\times$ 10$^{32}$ ergs s$^{-1}$
pc$^{-3}$

In the Orion Nebula (Garmire et al. 2000), one of the densest PMS
clusters known, the luminosity of X-ray emitting stars in 1 pc$^3$
is $\sim 10^{32}$ ergs s$^{-1}$.  Either the Carina region has a 
population of young stars denser than in most pre-main sequence 
regions or much of this emission is truely diffuse.  
The X-ray spectra of these regions are thermal.  If the regions have 
depth approximately equal to angular extent, and the emission is due
to diffuse, hot gas, over half the X-ray emission from the 
Carina Nebula originates in the diffuse component.
A gas density of $\sim$0.5 particles cm$^{-3}$
is required.  O star winds can easily supply the energy.

\section{Upper Limits to Undetected O and B Stars}

As is well known, B stars in general do not produce X-rays.
Because of the extensive optical work that has been done on the 
 stars in the two clusters, they provide a good place to explore 
the exact X-ray boundary.  To do this, we have derived upper limits 
to the X-ray flux of the hot stars in Table 4 of MJ93 which 
are not detected in X-rays.  
Fig 14  shows  the locations of these stars overlaid on 
the X-ray image.  In Tr 16 in the center of the image, the circled
locations appear to be free of sources.  In Tr 14 in the upper 
right, the situation is  more complex.  
 Tr 14 is a crowded region, and  sources
 in that region may have source confusion 
because of the large PSF  off-axis.    

We have adopted the criterion for source detection (Section 2)
from Gehrels (1986).  It is parameterized as a function of the number of 
background counts, which is a function of the off-axis distance.  
We have compared the measured counts  at the positions of the
O and B stars in MJ94 (Table 4) which are not in Table 1  
 with the detection level predicted by the Gehrels criterion.
Counts were measured in 2$\farcs$5 circles centered on these 
positions.  For the center of the field, this is a generous 
allowance for the PSF (0$\farcs$68) plus an
uncertainty in the pointing of 1".  Figure 15 shows that
 the measured 
counts are typically well below the detection criterion.
(Note that the counts from which sources were selected for Tables 
1-4 were based on {\tt WAVDETECT} results;  those in Fig 13 are from 
extracted counts with no background subtraction.)  The detection 
criterion used by Green et al. (1995) for a QSO survey is 
C$_{src} \geq 4 \sqrt{C_{bkg}}$, where C$_{src}$ and C$_{bkg}$
are the counts in the source and background respectively.  
For the very low background in the center of a 1 ksec exposure,
as seen in Fig 15, 
one background count corresponds to a detection limit of 4 source 
counts.  As summarized in Fig 15, the Gehrels criterion provides 
a conservative upper limit for sources concealed in the measured counts.  

Further from the axis, the PSF, and hence the 
included background counts, increase.
For comparison with the detection criterion, counts for locations 
greater than 7' off axis were measured in circles of 5" and 10"
radius as well as 2$\farcs$5 (not shown in Fig 15).  
For radii less than 10' off axis, the PSF is less than 10".
Counts in the 10" circles for several of the off-axis locations are above the 
detection criterion.  As discussed below, there are probably 
actually sources at several of these locations.  We note also 
that the Tr 14 region is a dense and complex one, and it is highly
likely that additional sources are included as the measurement area
is increased.  Because of this complexity, further discussion of
the upper limits far off axis does not seem warranted.

With these explorations in mind, we adopt the Gehrels limit as the 
upper limit of counts at the location of early B stars.

The measured counts at the location of O and B stars from 
MJ93 Table 4 (which are not detected sources)
is listed in Table 7.  For each
location, Table 7 provides the number (independent from the numbering
in Tables 1 to 4), the position,  the counts in the source and
background areas, and the net counts, all with their errors.  
Counts are measured in a circle with a 5" radius; background counts are
measured in an annulus between 5" and 10" radii.  Table 7 
lists the radius from the aim-point for each source and the 
spectral type (where available from MJ93). 
Spectral types are abbreviated to the temperature portion 
for plotting purposes. Upper 
limits to the luminosity (in ergs sec$^{-1}$) are computed 
for the Gehrels criterion 
 and the distance to the clusters.  

Fig 16 shows the {\it net} counts at the location of hot stars.
Counts were measured in a 2$\farcs$5 circle around the stellar positions.
Background counts were measured an annulus between 5" and 10". 
 The scatter around measures for Tr 16 near the center of the field is
well balanced around zero, confirming that these are non-detections. For stars
further off axis, the 2$\farcs$5 circle, of course, does not include the full
PSF, and hence some source counts will be included in the region which defines
the ``background" counts. Even though the measures will be ``over-corrected"
for background, the Tr 14 stars have predominately positive source counts.
This indicates that the central concentration in the PSF is such that a number
of Tr 14 stars in fact are probably X-ray sources, even though they do not
pass the detection criterion.

Fig 17 shows the upper limit luminosity from Table 6 for 
hot stars not detected in X-rays as a function of spectral type.
The two O stars which were not detected have comparatively high
upper limits because they are at the edge of the image.  The counts
corresponding to these upper limits are at or slightly above
 the lower envelope of the O star detections in Fig 5.
 Had they been at the image center, they might have been detected.
  On the
 other hand, the B stars from the center of the image have a typical count
  upper limit (log Lum = 30.8 ergs sec$^{-1}$) which is below the 
 counts of any detected hot star in Fig 8.

We can use Fig 4
 to examine the dividing line between detected X-rays and 
the lack of X-rays.  With the exceptions noted above, all O stars
are detected.  For the B0 spectral types, two stars are detected and one
is not.  Cooler B stars are not detected with one exception.  This 
points to a very clean divide in spectral type at B0.  
It also provides us with a list of 
14 stars in Tr16 for which we can constrain the properties 
of a cool companion star.  This upper limit only overlaps with 
the weakest sources in the distribution of X-ray luminosities 
(Fig 9).  While we would not yet have detected 
a companion comparable to the faintest 
X-ray sources among late type cluster members, we can say that 
the B stars do not have  X-ray bright companions.

\section{Hardness Ratios}

The hardness ratios of X-ray sources provide the simplest 
description of their spectra.  In this section, we explore 
the hardness ratios  (Tables 3 and 4) as defined in Section 2
for both hot stars and other X-ray sources.  We limit the 
sources examined to the brightest so 
 the errors in the 
ratios are small enough so that they do not obscure any trends
or comparisons.  We have examined the distribution of errors in 
the hardness ratios as a function of luminosity.
 For the O-B stars (Table 3),  if we use only sources with 
 log L$_X$ $\ge$ 32.2 ergs sec$^{-1}$, the uncertainty in the 
 soft hardness ratio, HR$_{MS}$ is less than or equal 
 to 0.1 (with one exception);
 for the hard hardness ratio, HR$_{HM}$, the error is less than
 or equal to 0.15. The other stars (Table 4, referred to 
 as cool stars below) are somewhat weaker in X-rays than the 
 hot stars, so we use sources with 
 log Lum $\ge$ 32.0 ergs sec$^{-1}$.  This results 
 in uncertainties less than 0.2 in both colors (after two sources
 with large errors were removed from the sample).
Our sample is thus limited to only the brightest sources. 

Fig 18 shows the resulting hardness ratio plot. 
The hardness ratio created from the soft and medium bands (0.5 to 0.9 keV, 0.9
to 1.5 keV) can be either positive or negative, i.e., either band can have
more counts whereas the 
hardness ratio from the medium and hard band (1.5 to 2.04
keV) is always negative, i.e. the hard band always has fewer counts. 
The hot stars
fall in a diagonal sequence with a scatter consistent with the uncertainties
in the colors. There are only two cool stars with well determined 
hardness ratios, which are both are very hard in Fig. 18.

We have explored this locus further by creating  hardness ratios
in the same bands using PIMMS (v3.2).  
The basic model (solid line in Fig 18) is 
a Raymond-Smith model with solar abundance and  a temperature
range from log T = 6.0 to 7.2.  A neutral column density 
N$_H$ = 3 x 10$^{21}$ cm $^{-2}$ was used corresponding to the 
typical extinction for the field E(B-V) = 0.52 mag.  For two 
temperatures, N$_H$ was increased to  5 x 10$^{21}$ cm $^{-2}$,
corresponding to E(B-V) = 0.9 mag, an extreme value found for a 
few stars in the field.  The result of this difference in 
extinction is shown in Fig 18.  
For comparison, we have computed hardness ratios for 0.2 solar
abundance (dashed line in Fig. 18).  We have also computed models 
for thermal bremsstrahlung (kT from 0.2 to 1.0 keV).
Fig 18 shows that a range of HRS (the soft ratios) can be obtained with 
any of the models.  However,  
the observed ratios for both hot and cool 
stars lie systematically at larger (harder) values of HRH than 
the computed values for either Raymond-Smith plasma or 
bremsstrahlung.  Although the bremsstrahlung ratios for the 
hottest temperature match the ratios for the hot stars, for cooler 
temperatures they are a 
poorer match than the Raymond-Smith models.  Based on these models, 
the HRS values for the hot stars indicate a range of temperatures
from approximately 3 x 10$^6$ to more than 10$^7$.

How are these X-ray hardness ratios related to the  optical (photospheric)
properties of the stars?  In Fig 18, the 3 hot stars with negative
HR$_{MS}$ are HDE 303308 [E(B-V) = 0.46 mag; O3 V], HD93205
 [E(B-V) = 0.40 mag; O3 V], and -59$^o$2600  [E(B-V) = 0.51 mag; O6 V].
In other words, all three have reddening typical of the field. 
The stars with HR$_{MS}$ near 0.1 are HD 93250 [E(B-V) = 0.49 mag; O3 V]
and HD 93160 [E(B-V) = 0.31 mag; O6 III].
  Interpreting
the hardness ratios in terms of photospheric temperature arrives at the
contradiction that the X-ray temperatures
are hotter than the  O3 V stars in the first group, although 
the photospheric temperatures are similar or cooler.
The four hot stars with distinctly harder  HR$_{MS}$ are 
Tr 16-22 (O8.5 V, E(B-V) = 0.78 mag), Tr 16-244 (O3 I, E(B-V) = 0.94 mag),  
HD 93129AB (O3 I, E(B-V) = 0.55 mag), HD 93162 = WR 25 (WN7 + abs,
E(B-V) = 0.66 mag).  None of these are normal main sequence objects.
HD 93129AB and Tr 16-22 are both among the hottest stars known
(O3), and also very luminous.  The fact that HD 93129AB apparently
has the normal reddening for  Tr 14 suggests that  the very hard
HRH ratio does not result from circumstellar material. This is in contrast to  
Tr 16-244 which has the same temperature and luminosity, but is more 
heavily reddened that the average cluster member.  Both HD 93162
and Tr 16-22 stand out in Fig. 10 because they have a particularly
large L$_X$/L$_{bol}$.  We already know that HD 93162 is unusual, 
since not only has it evolved to be a very early
stage in the  Wolf-Rayet sequence, it 
is also unusually X-ray luminous for a  Wolf-Rayet star.  Tr 16-22,
on the other hand, is cooler and fainter than stars evolving off
the main sequence in the clusters.  Its unusually hard and 
bright X-ray luminosity must be linked with some unusual 
property, such as a massive binary companion.  

The two cool stars with well-determined hardness ratios are 
both among the hardest stars in the field.  Interestingly, one is located
within the range of the cool pre-main sequence stars in Fig. 7, but one 
is one of the X-ray bright A stars.  

Fig 19 confirms that the HRS (the soft hardness ratio) 
is not correlated with photospheric temperature indicated
by spectral type.  The O3 stars, in particular, 
cover a very large range in 
HRS.  

How do we interpret the hardness ratios of the X-ray sources in Tr 14 and Tr
16? There seem to be two important considerations. First, we are looking at a
biased sample. Because we have restricted the sample to sources with
reasonably well determined ratios, we are only looking at the most luminous
sources. Second, both the hot and cool stars have more counts in the hard band
than predicted from the simple PIMMS calculations. A plausible interpretation
for this is that we have used a one temperature model in the calculations, but
the sources may actually have a second harder component. This is frequently
seen in spectral fits of stellar sources. We note that the discussion of the
individual sources above rules out single parameter explanations of the
hardness based on photospheric temperatures or circumstellar reddening. In
summary, unusually X-ray bright sources are surprisingly hard.

\section{Discussion}

\subsection{Binaries}

We have assembled the X-ray properties of sources in Tr 14 and Tr 16,  
  including several sources which are unusually strong or 
hard in X-rays.  We have already discussed properties such as 
reddening, spectral type and luminosity in relation to 
the X-ray properties.  One further property (at least) may be important in 
X-ray production, namely whether the star is single, a member of a 
close binary system, or a member of a wide binary system.  

  In O stars,  a  binary companion
with sufficient mass to drive a wind of its own may
enhance X-ray production in via a ``colliding wind". 
  A good 
deal of work has been done  to establish whether 
the stars in the Carina clusters are single or multiple.
  The most definitive study, 
 with the highest resolution spectra, is that of Penny et al. (1993).
They concluded that none of the six bright members of Tr 14 are  
short-period spectroscopic binaries.  
This is in contrast to an earlier 
study at lower resolution by Levato et al. (1991b).  However, the 
Penny work, and a later study by Garcia et al. (1998), demonstrate 
that Tr 14 is surprisingly deficient in short-period
binaries among the hottest
stars.  Penny et al. suggest that the cluster has remained compact 
because the lack of binaries lowers the cluster heating via binary-binary 
interactions, although the results come from, of course, only the 
brightest stars.

Velocity studies of 26 stars in 
Tr 16 were carried out by Levato et al. 
(1991a), using  the same equipment 
 used for the  Tr 14 study, with relatively low resolution
(43 \AA\/ mm$^{-1}$). They found  velocity variation in at least 
50\% of the stars studied.   The work by this team in 
Tr 14 was superceded by higher resolution work, 
and the radial velocity variations they found in Tr 16 need to be 
confirmed.  It would be quite surprising 
to have dramatically different binary fractions in the  two clusters. 
  Three of the 
X-ray sources in the ACIS field have recently been observed by the 
X-Mega campaign to obtain optical spectroscopy of the hot stars
which are X-ray sources.  Colombo et al. (2001) have observed 
CPD -59$^0$2635.  They found that it is a double-line spectroscopic
binary and determined an orbit.  Both components are massive and 
Colombo et al. determine spectral types of O8 V and O9.5 V.
HD 93205 has been observed by Morrell et al. (2001), who  
report that it is 
another double-lined spectroscopic binary with O3 V and O8 V 
components.  CPD -59$^0$2636 is a triple system, with 
O7 V, O8V, and O9 V components (Albacete Colombo et al. 2002).
Components A and B are a binary with a period of 3.6 days. 
Component C is itself a binary with a period of 5.0 days.
CPD -59$^0$2603 = Tr 16-104 has been observed by Rauw et al. 2001.
They find it to be a triple system with O7 V, O9 V, and B0.2 IV
components.  The close binary (O7 V and O9 V stars) has a period
of 2.2 days, and short eclipses.

Recently Nelan et al. (2002, private communication) have resolved 
HD 93129 A with the Hubble Space Telescope FGS (fine guidance system).
This may affect the X-ray properties, although the two 
components are much more widely separated than the binaries
in Table 8, so it is not clear that it does.

We have collected radial velocity 
information about the single/binary nature of the 
X-ray sources in the field in Table 8.
 It shows the hot stars 
within the Chandra field, the source of velocity information,
the binary status, and the orbital period where available.  
 Mason et al. (1998)
provide a convenient summary of velocity work on a number of these
stars and also speckle interferometry to sample wider binaries.  
In Table 8 we have marked the velocity variations from Levato et al. (1991a)
with  ``:" to indicate that they need confirmation. 

Identifying multiple systems in hot stars with broad lines is not 
easy.  In addition to the radial velocity studies, as discussed above,
van der Hucht (2001) has suggested that the Wolf-Rayet star HD 93162 
may be a binary on the basis of diluted emission lines.  Penny (1996) 
has suggested that HD 93250 might be a binary on the basis of variable
line widths.  This is in accord with the suggestion
of a companion (Walborn, 
Nichols-Bohlin, and Panek, 1985) based on unsaturated C IV line 
cores. Confirmation of the single/multiple status of HD 93250 
is a very important question in the interpretation of the 
X-ray flux of this extremely luminous object.

Table 8 can be used to
 compare what is known about the single/binary
nature of X-ray sources that are unusually strong, using, for 
instance, L$_X$/L$_{bol}$ in Table 6. For the Wolf-Rayet 
star HD 93162, L$_X$ is particularly strong, but evidence for 
a binary system is only indirect.  
 Among the O3 stars, HD 93250 is a strong, and also hard 
 X-ray source, and again has only indirect evidence of a possible
 binary companion.  HD 93129AB is a strong and also hard X-ray source,
 although its L$_X$/L$_{bol}$ is not remarkable.  There is no evidence for
 a close binary companion.  Tr 16-22 is a remarkably strong X-ray source,
 but does not have radial velocity 
 studies.  On the other hand, HD 93205,
 which is a double lined spectroscopic binary (Morrell et al. 2001)
 with an orbital period of
 6.1 days and spectral types of O3 V and O8 V, is not a particularly
 strong X-ray source.   CPD -59$^o$ 2635 is also a double-lined 
 spectroscopic binary (Colombo et al. 2001), with a period of 2.3 days
and spectral types O8 V and O9.5 V, and again is not  a particularly
strong X-ray source.  CPD -59$^o$ 2636 is a similar case of 
a multiple system with two short period binaries (Albacete Colombo et al.
2002), but an unremarkable X-ray flux.
In summary, the two with particularly hard
spectra (HD 93129AB and HD 93250) may well not be binaries.  On the 
other hand three sources in short period binaries made up of  
O star components (HD 93205, CPD  -59$^o$ 2635, and CPD  -59$^o$ 2636)
 are not particularly remarkable X-ray sources.

The emerging picture is that while close binaries with two hot 
stars may produce a strong X-ray flux (the colliding wind
model),  this does not seem to be the whole 
explanation for surprisingly
strong X-ray sources.  Clearly there is the standard problem that 
there is almost never enough information to prove that a star is 
single.  However,  it is short period binaries with 
massive companions that  should be   colliding 
wind systems, and they  are the easiest to find.  With the
incomplete velocity evidence that is available, strong sources 
and binary orbits are nearly anti-correlated.   We suspect this 
demonstrates that our knowledge of binary systems is 
incomplete.

 HD 93162 = WR 25, and HD 93250 are surprisingly strong in X-rays,
 but with only indirect suggestions of a binary companion.  Should
 it turn out that the X-ray strength does come from a companion, 
 this would provide an important way to get  more complete statistics on
 the binary fraction in massive stars, since such strong X-ray 
 sources are not overlooked.  

Chlebowski and Garmany (1991) provide comparisons of X-ray flux and 
stellar parameters such as wind velocity and luminosity for an 
extensive list of O stars.  They find that in general binaries
do not produce larger mass loss than single stars.  On the other hand, 
close binaries seem to produce more X-ray flux than either contact
binaries or wider binaries.  One way to reconcile these trends with 
the apparent lack of correlation between binary status in Table 8 and 
excess X-ray flux is that we are sampling regimes where different 
effects become dominant.  This is particularly likely for the 
group of extreme stars found in the Carina nebula: the most massive, 
the most luminous.  For instance, for the most massive stars, 
extreme mass loss rates may dominate over binary-single trends and 
produce an unusually large X-ray flux.  For stars less massive than
O3 stars, the effect of a close binary with a massive companion 
may become evident in the X-ray flux.

\subsection{HD 93250}

As summarized above, there is no {\it velocity} evidence the  HD 93250
is a short period binary system, although there are suspicions that
it might have a companion from line-width variations and line depths.
However the recent reexamination by Walborn et al. stress line depth
in the context of luminosity not companions.  This paper stresses that 
complications in the interpretation of these extreme stars because of 
binary companions must be understood using Milky Way examples.  Even 
though there are many examples in the Magellanic clouds, 
observations are
typically contaminated by a surrounding cluster.   If HD 93250 is 
single (and there is no clear evidence at present against this), it 
is a pivotal object in understanding the X-ray production and 
evolution of very massive stars.  It is an extremely strong (and also
hard) X-ray source in comparison with other O3 stars. 
Among the Tr 14 and Tr 16 stars it has a V magnitude second only to 
HD  93129AB (MJ93). The hardness ratio HRS,
however, lies between the softer O3 stars and the harder but unusual
and more luminous HD 93129AB and Tr 16-244.  If X-ray flux of HD 93250 is 
not affected by a companion, it provides a benchmark 
of an extremely luminous main sequence 
dwarf with a very high L$_X$/L$_{bol}$
produced by the outer atmosphere processes of the star itself.  
Since it will presumably accelerate its mass loss as it evolves off
the main sequence, it provides an important example of high X-ray production
of a star before its mass loss becomes extreme.

\subsection{B Stars }

There is another regime in which X-ray observations provide 
an important handle on the binary fraction in massive stars.  
The upper limits to X-rays from B stars (Fig. 17) apply also 
to the X-ray flux from a cool companion.  
For instance, the  upper limit to the luminosity 
(log L$_X$ = 30.8 ergs sec${^-1}$) 
would pick up many of the cool star 
  X-ray sources in the Orion Nebula cluster 
  (Feigelson, et al. 2002).
In other words, a deeper image could rule out most cool companions.  
Quantifying how what fraction of companions would be detected
at a given sensitivity level,
however, is not easy.  We can estimate from Fig 10 of 
Feigelson, et al. that less than a quarter of the Orion cool
stars would be detected with the sensitivity limit of our $\eta$
Car image (including a small approximate correction to account 
for absorption in Orion).  However, Fig. 12 in Feigelson, et al.
impiles that companions of A and late B stars have a higher 
luminosity than the average cool star.  That is, for a given 
sensitivity limit, a larger fraction of companions would be 
detected than estimated from the luminosity distribution of 
late type stars.  
Accurately identifying binary companions much less massive than 
the primary is very difficult to do, particularly for massive 
B stars with only a few broad lines to measure velocities from. 
X-ray studies provide detection of this otherwise elusive 
population of companions.

\subsection{Colliding Wind Binaries}

 The emerging picture of colliding wind binaries is that they 
include systems with  widely separated components, as 
well as systems with close components.  WR 140 with 
an orbital period of 8 years is an example. 
Our summary of velocity information about binaries in Table 8,
and the discussion in section 10.1 indicate that the 
relation between binary status and strong X-rays is 
not simple.  The recent XMM-Newton spectrum of the 
Wolf Rayet star WR 110 (Skinner et al. 2002) suggests 
yet another complication in interpretation.  They find  a 
cool soft component to the spectrum, as expected from a 
wind-shock model, and  in addition, they find a hard X-ray component.
They examine a number of possibilities for producing this
component and  conclude that it can be produced by a wind 
and a close but as yet undetected companion.  However, if the 
companion is not found, they suggest the possibility that the 
hard component is produced by the Wolf Rayet star itself.




\section{Summary}

The Chandra image of $\eta$ Car and the clusters Tr 14 and Tr 16 
has provided a number of new results.

$\bullet$  Although the hottest stars have been known
to be X-ray sources since the Einstein Observatory observations, 
 this is the first time the population of 
X-ray sources associated with cooler stars has been discussed.

$\bullet$ Two new anomalously strong X-ray sources have 
been found among the hot star population.  
 Tr 16-244, a heavily reddened O3 I star, and Tr 16-22, a 
 heavily reddened O8.5 V star have surprisingly large X-ray 
 fluxes.  They are both close to the dust lanes which cross 
 the sides of the cluster and which may be responsible for the 
 above average reddening.
 
$\bullet$  Several stars have an unusually large L$_X$/L$_{bol}$.
HD 93162, a Wolf-Rayet star, has long been known to have an even 
higher L$_X$/L$_{bol}$ than   other Wolf-Rayet stars. 
There is some indication that it may be a binary, which could account
for this.   Tr 16-22
also stands out in L$_X$/L$_{bol}$, perhaps indicating a
colliding wind binary, 
although there are no radial velocity studies to 
provide further information. The third star with a large 
L$_X$/L$_{bol}$ is Tr16-11, the B1.5 star.   
The X-ray flux may be produced by a second cool star.    

$\bullet$ We confirm that the OB stars are on average more luminous than the 
other X-ray sources which are  pre-main sequence stars,
as has been found in other regions.  


$\bullet$ The X-ray sources (cool stars) define a locus in the color-magnitude
diagram  which lies above stars which are presumably 
field stars in the Carina arm.  

$\bullet$  We have put upper limits on the OB stars in the 
field which we do not detect. For comparison, in this 
short exposure, if companions to the B stars had an X-ray luminosity 
distribution like the late type stars in the Orion nebula cluster, 
we would  detect only a fraction of  the companions.

$\bullet$ We discuss the hardness ratios of the brightest 
sources in bands 0.5 to 0.9
keV, 0.9 to 1.5 keV, and 1.5 to 2.04 keV, though the sample is limited to only
the brightest sources. 
Hardness ratios are not correlated in a simple way with 
photospheric temperature (spectral type).
Both hot and cool stars lie parallel to a 
locus derived for an isothermal plasma, however 
there is an offset indicating extra counts in the highest energy band.
This could be due to a second component at a higher temperature, at least for
these very luminous sources.

$\bullet$ We summarize the information available about the binary 
orbital velocity motion of the O star X-ray sources.

Acknowledgments:  It is a pleasure to thank everyone who 
worked on the Chandra project for making this data possible.
We thank Nuria Huelamo (MPE) for obtaining 
 optical images of the field.  Useful comments were 
provided by D. Gies. We thank an anonymous referee for 
for comments which improved the presentation of the 
results. 
Financial assistance was provided from the 
Chandra X-ray  Center NASA Contract NAS8-39073

\clearpage

\begin{deluxetable}{crrrrrrrrrrr}
\footnotesize
\tabletypesize{\scriptsize}
\tablecaption{ X-Ray Sources: OB stars
 \label{tbl-1}}   
\tablewidth{7.0truein}
\tablehead{
\colhead{Src} & 
\colhead{RA} & \colhead{ Dec} & \colhead{Rad} & \colhead{Total}& \colhead{Err}
 & \colhead{Soft}& \colhead{Err}
 & \colhead{Med}& \colhead{Err}
 & \colhead{Hard}& \colhead{Err}   \\
\colhead{} & 
\colhead{h m s} & \colhead{ $^o$ ' "} & \colhead{'} & \colhead{counts}& \colhead{counts}
 & \colhead{counts}& \colhead{counts}
 & \colhead{counts}& \colhead{counts}
 & \colhead{counts}& \colhead{counts}   \\
  }

\startdata

   2 &  10 45 16.55 & -59 43 36.9 &  3.0 &   84.7 &    9.1 &   43.5 &    6.6 &   33.9 &    5.7 &    7.3 &    2.6  \\ 
   5 &  10 45 12.89 & -59 44 19.3 &  3.5 &  114.0 &   10.5 &   50.7 &    7.1 &   56.1 &    7.4 &    7.2 &    2.6  \\ 
   8 &  10 45  8.23 & -59 40 49.2 &  0.6 &   32.6 &    5.7 &   12.6 &    3.6 &   20.0 &    4.5 &    0.0 &    0.0  \\ 
   9 &  10 45  6.74 & -59 41 56.4 &  1.0 &   15.4 &    4.0 &    9.8 &    3.2 &    5.8 &    2.5 &   -0.2 &    0.1  \\ 
  10 &  10 45  5.90 & -59 40  5.8 &  1.0 &  232.1 &   15.3 &  141.4 &   11.9 &   73.7 &    8.6 &   17.0 &    4.1  \\ 
  22 &  10 44 45.04 & -59 33 54.0 &  7.5 & 1572.8 &   37.2 &  597.6 &   23.0 &  771.7 &   26.0 &  203.5 &   13.4  \\ 
  25 &  10 44 33.77 & -59 44 15.6 &  4.9 &  532.6 &   22.5 &  311.4 &   17.2 &  189.9 &   13.4 &   31.4 &    5.5  \\ 
  28 &  10 44 10.38 & -59 43 11.3 &  7.0 & 4989.9 &   67.3 &  733.2 &   25.9 & 3018.1 &   52.3 & 1238.5 &   33.5  \\
\enddata

\end{deluxetable}                                                               
  
\clearpage

\begin{deluxetable}{crrrrrrrrrrr}
\footnotesize
\tabletypesize{\scriptsize}
\tablecaption{ X-Ray Sources: Not OB Stars
 \label{tbl-2}}   
\tablewidth{7.0truein}
\tablehead{
\colhead{Src} & 
\colhead{RA} & \colhead{ Dec} & \colhead{Rad} & \colhead{Total}& \colhead{Err}
 & \colhead{Soft}& \colhead{Err}
 & \colhead{Med}& \colhead{Err}
 & \colhead{Hard}& \colhead{Err}   \\
\colhead{} & 
\colhead{h m s} & \colhead{ $^o$ ' "} & \colhead{'} & \colhead{counts}& \colhead{counts}
 & \colhead{counts}& \colhead{counts}
 & \colhead{counts}& \colhead{counts}
 & \colhead{counts}& \colhead{counts}   \\
  }

\startdata

   1 &  10 45 22.01 & -59 40 52.5 &  2.3 &   23.7 &    4.9 &    3.9 &    2.0 &   15.8 &    4.0 &    4.0 &    2.0  \\ 
    3 &  10 45 16.25 & -59 41 41.0 &  1.7 &   18.1 &    4.2 &    3.0 &    1.7 &   13.1 &    3.6 &    2.0 &    1.4  \\ 
   4 &  10 45 14.80 & -59 40  7.1 &  1.7 &   12.9 &    3.6 &    3.9 &    2.0 &    6.9 &    2.6 &    2.0 &    1.4  \\ 
   6 &  10 45 11.79 & -59 41 33.8 &  1.2 &   27.7 &    5.3 &    6.9 &    2.6 &   14.8 &    3.9 &    6.0 &    2.5  \\ 
   7 &  10 45  8.57 & -59 40 16.7 &  1.0 &   27.9 &    4.9 &    8.1 &    2.6 &   16.3 &    3.7 &    3.5 &    1.7  \\ 
  11 &  10 45  4.21 & -59 40 10.7 &  0.9 &   15.9 &    4.0 &    4.9 &    2.2 &    9.0 &    3.0 &    2.0 &    1.4  \\ 
  12 &  10 45  3.89 & -59 38 19.9 &  2.7 &   20.9 &    4.5 &    4.2 &    2.0 &   13.6 &    3.6 &    3.1 &    1.7  \\ 
  14 &  10 45  2.93 & -59 42 25.9 &  1.4 &   13.8 &    3.7 &    1.8 &    1.4 &   11.1 &    3.3 &    0.9 &    1.0  \\
\enddata

\end{deluxetable}

\clearpage

\begin{deluxetable}{crrrrccrr}
\footnotesize
\tabletypesize{\scriptsize}
\tablecaption{ X-Ray Sources: Hardness Ratios: OB Stars 
 \label{tbl-3}}   
\tablewidth{6.5truein}
\tablehead{
\colhead{Src} & 
\colhead{HR$_{MS}$} & \colhead{Err} & \colhead{HR$_{HM}$} & \colhead{Err}
& \colhead{log Flux}
 & \colhead{Err}& \colhead{log Lum}
 & \colhead{Err}   \\
\colhead{} & 
\colhead{} & \colhead{} & \colhead{} & \colhead{}
& \colhead{ergs cm$^{-2}$ s$^{-1}$}
 & \colhead{ergs cm$^{-2}$ s$^{-1}$ }& \colhead{ergs s$^{-1}$}
 & \colhead{ergs s$^{-1}$}   \\  }

\startdata

   2  &  -0.12  &   0.14  &  -0.65  &   0.20  & -12.91  & -13.88  &  31.96 &  30.99 \nl 
   5  &   0.05  &   0.12  &  -0.77  &   0.17  & -12.78  & -13.82  &  32.09 &  31.06 \nl 
   8  &   0.23  &   0.22  &  -1.00  &   ----  & -13.33  & -14.08  &  31.55 &  30.79 \nl 
   9  &  -0.26  &   0.32  &  -1.07  &   0.62  & -13.65  & -14.24  &  31.22 &  30.64 \nl 
  10  &  -0.31  &   0.08  &  -0.63  &   0.14  & -12.47  & -13.66  &  32.40 &  31.22 \nl 
  22  &   0.13  &   0.03  &  -0.58  &   0.04  & -11.64  & -13.27  &  33.23 &  31.61 \nl 
  25  &  -0.24  &   0.05  &  -0.72  &   0.09  & -12.11  & -13.49  &  32.76 &  31.39 \nl 
  28  &   0.61  &   0.02  &  -0.42  &   0.02  & -11.14  & -13.01  &  33.73 &  31.86 \nl 
  39  &   0.14  &   0.28  &  -0.84  &   0.39  & -13.54  & -14.20  &  31.33 &  30.67 \nl 
  44  &  -0.06  &   0.16  &  -0.72  &   0.23  & -13.05  & -13.96  &  31.82 &  30.91 \nl 
  45  &  -0.03  &   0.16  &  -0.64  &   0.22  & -13.03  & -13.94  &  31.84 &  30.93 \nl 
  55  &   0.51  &   0.10  &  -0.43  &   0.09  & -12.49  & -13.67  &  32.39 &  31.20 \nl 
  58  &  -0.11  &   0.24  &  -1.00  &   ----  & -13.41  & -14.12  &  31.47 &  30.75 \nl 
  59  &  -0.18  &   0.30  &  -0.31  &   0.38  & -13.50  & -14.15  &  31.37 &  30.73 \nl 
  83  &  -0.54  &   0.18  &  -0.46  &   0.32  & -13.12  & -13.98  &  31.75 &  30.89 \nl 
  88  &  -0.18  &   0.09  &  -0.82  &   0.15  & -12.59  & -13.71  &  32.29 &  31.17 \nl 
  96  &  -0.49  &   0.12  &  -0.94  &   0.26  & -12.80  & -13.84  &  32.07 &  31.04 \nl 
 102  &   0.66  &   0.13  &  -0.37  &   0.11  & -12.69  & -13.75  &  32.19 &  31.12 \nl 
 103  &   0.45  &   0.03  &  -0.49  &   0.03  & -11.51  & -13.22  &  33.36 &  31.65 \nl 
 135  &   0.02  &   0.27  &  -0.61  &   0.35  & -13.43  & -14.11  &  31.45 &  30.77 \nl 
 141  &   0.75  &   0.53  &  -0.41  &   0.39  & -13.72  & -14.16  &  31.16 &  30.71 \nl 
 145  &   0.16  &   0.07  &  -0.66 &   0.08 & -12.23 & -13.52 &  32.65 &  31.36 \\
\enddata                                                                        
                                                                                
\end{deluxetable}

\clearpage

\begin{deluxetable}{crrrrccrr}
\footnotesize
\tabletypesize{\scriptsize}
\tablecaption{ X-Ray Sources: Hardness Ratios: Not OB stars
 \label{tbl-4}}   
\tablewidth{6.5truein}
\tablehead{
\colhead{Src} & 
\colhead{HR$_{MS}$} & \colhead{Err} & \colhead{HR$_{HM}$} & \colhead{Err}
& \colhead{log Flux}
 & \colhead{Err}& \colhead{log Lum}
 & \colhead{Err}   \\
\colhead{} & 
\colhead{} & \colhead{} & \colhead{} & \colhead{}
& \colhead{ergs cm$^{-2}$ s$^{-1}$}
 & \colhead{ergs cm$^{-2}$ s$^{-1}$ }& \colhead{ergs s$^{-1}$}
 & \colhead{ergs s$^{-1}$}   \\
  }

\startdata

   1 &   0.60 &   0.29 &  -0.60 &   0.29 & -13.47 & -14.15 &  31.41 &  30.73 \\ 
   3 &   0.62 &   0.32 &  -0.73 &   0.34 & -13.58 & -14.21 &  31.29 &  30.66 \\ 
   4 &   0.28 &   0.38 &  -0.55 &   0.43 & -13.73 & -14.28 &  31.14 &  30.59 \\ 
   6 &   0.36 &   0.27 &  -0.43 &   0.28 & -13.40 & -14.12 &  31.48 &  30.76 \\ 
   7 &   0.34 &   0.24 &  -0.64 &   0.28 & -13.39 & -14.15 &  31.48 &  30.72 \\ 
  11 &   0.29 &   0.33 &  -0.64 &   0.39 & -13.64 & -14.24 &  31.24 &  30.64 \\ 
  12 &   0.53 &   0.30 &  -0.63 &   0.31 & -13.52 & -14.19 &  31.36 &  30.69 \\ 
  14 &   0.73 &   0.37 &  -0.84 &   0.39 & -13.70 & -14.27 &  31.17 &  30.61 \\
\enddata

\end{deluxetable}

\clearpage






\begin{deluxetable}{lrrrrrrr}
\tabletypesize{\scriptsize}
\tablewidth{0pt}
\tablecaption{Photometry of Low Mass X--ray sources}
\tablehead{
\colhead{X-Ray ID}           & \colhead{RA}      &
\colhead{Dec}          & \colhead{B-V}  &
\colhead{V}      &
\colhead{V-I}  & \colhead{Notes}  
}
\startdata
1	& 10:45:22.01  &	-59:40:52.5	&1.589	&16.914	&1.930&      \\ 
3	& 10:45:16.25  &	-59:41:41.0	&0.421	&13.377 &0.770&	   \\
4	& 10:45:14.80  &	-59:40:07.1	&-- --	&18.452	&1.734&	   \\
6	& 10:45:11.79  &	-59:41:33.8	&2.009	&19.178	&2.054&	   \\
7	& 10:45:08.57   &	-59:40:16.7	&1.106	&15.410	&1.350&	   \\
11	& 10:45:04.21   &	-59:40:10.7	&-- --	&-- --	&-- --&	 3 \\
12	& 10:45:03.89   &	-59:38:19.9	&-- --	&-- --	&-- --&	 3 \\
14	& 10:45:02.93   &	-59:42:25.9	&1.759	&17.852	&2.058&	 3 \\
\enddata

\tablenotetext{1}{Two  stars}
\tablenotetext{1a}{Not resolved; not plotted}
\tablenotetext{1b}{ Not resolved; estimated combined magnitude}
\tablenotetext{1c}{Resolved; position match poor; not plotted}
\tablenotetext{1d}{Resolved; likely source used}
\tablenotetext{2}{High background.}
\tablenotetext{3}{Contamination from nearby bright star}
\tablenotetext{4}{Uncertain identification}
\tablenotetext{5}{Faint star}
\tablenotetext{6}{The star is overexposed. V$<$12.5}
\tablenotetext{7}{Outside the observed fields}


\end{deluxetable}

%
%
%

\clearpage

\begin{deluxetable}{lrrlcrcrrr}
\footnotesize
\tabletypesize{\scriptsize}
\tablecaption{ X-Ray Sources: Hot Star Identifications
 \label{tbl-5}}   
\tablewidth{0pt}
\tablehead{
\colhead{Source} & 
\colhead{Table 3} & \colhead{ V } & \colhead{ Spect. } 
& \colhead{E(B-V) } & \colhead{ M$_{bol}$ } & \colhead{log (L$_X$/L$_{bol}$) }   \\
\colhead{ } & 
\colhead{ID} & \colhead{ mag } & \colhead{ Type } 
 & \colhead{ mag }  & \colhead{mag }  & \colhead{  } \\
  }
\startdata

HD93129AB  &  103  &   6.90  & O3 I  &  0.55  &  -11.5 &  -6.72  \\
HD93250    &   22  &   7.41 &  O3 V  & 0.49   &  -10.7 &  -6.54 \\ 
HD93205    &   25  &  7.76   & O3 V  & 0.40   &  -10.1 &  -6.76  \\
HD93160    &  145  &  7.88   & O6 III & 0.31 &  -9.4  &  -6.60 \\
HD93162    &   28   &  8.11  & WN7 +abs   &  0.66 &  -10.1 &  -5.40 \\
HDE303308  &   10   &  8.19  & O3 V  & 0.46  &   -9.8 &  -7.00  \\
HD93204    &   96  &  8.48   & O5 V  & 0.41   &   -9.0 &  -7.02  \\
-59$^0$2600 &  88 &  8.65   & O6 V   & 0.51    &   -9.0 &  -6.80  \\
\enddata

  \end{deluxetable}                               
\clearpage

\begin{deluxetable}{crrrrrrrrrlr}
\footnotesize
\tabletypesize{\scriptsize}
\tablecaption{Non-Detections: OB stars
 \label{tbl-6}}   
\tablewidth{7.0truein}
\tablehead{
\colhead{Src} & 
\colhead{RA} & \colhead{ Dec}  & \colhead{Src}& \colhead{Err}
 & \colhead{Bkg}& \colhead{Err}
 & \colhead{Net}& \colhead{Err} & \colhead{Radius} & \colhead{ Spectral}  
 & \colhead{Log Lum}    \\
\colhead{} & 
\colhead{} & \colhead{ }  & \colhead{Cts}& \colhead{Cts}
 & \colhead{Cts}& \colhead{Cts}
 & \colhead{Cts}& \colhead{Cts} & \colhead{'} & \colhead{ Type}  
 & \colhead{Upper}    \\
 \colhead{} & 
\colhead{} & \colhead{ }  & \colhead{}& \colhead{}
 & \colhead{}& \colhead{}
 & \colhead{}& \colhead{} & \colhead{} & \colhead{}  
 & \colhead{Limit}    \\
 }

\startdata

   1 &  10 43 59.96 & -59 32 26.3 &   30.0 &    5.5 &   61.0 &    7.8 &    9.6 &    6.1 &   11.8 &    O6.5 &   31.613  \\ 
   2 &  10 44 26.54 & -59 41  02.8 &    3.0 &    1.7 &   15.0 &    3.9 &   -2.0 &    2.2 &    4.7 &   B1.5 &   30.925  \\ 
   3 &  10 44 38.73 & -59 48 14.3 &    6.0 &    2.4 &   14.0 &    3.7 &    1.3 &    2.7 &    7.8 &    ... &   31.208  \\ 
   4 &  10 44 56.27 & -59 33  04.0 &    2.0 &    1.4 &    7.0 &    2.6 &   -0.3 &    1.7 &    8.0 &   B0.5 &   31.228  \\ 
   5 &  10 44  00.97 & -59 35 46.2 &   12.0 &    3.5 &   13.0 &    3.6 &    7.7 &    3.7 &    9.5 &   O9 &   31.382  \\ 
   6 &  10 45  08.99 & -59 40 40.7 &    6.0 &    2.4 &   24.0 &    4.9 &   -2.0 &    2.9 &    0.8 &   ... &   30.813  \\ 
   7 &  10 44 40.38 & -59 41 49.0 &    2.0 &    1.4 &    6.0 &    2.4 &   -0.0 &    1.6 &    3.0 &   B1 &   30.843  \\ 
   8 &  10 45 11.22 & -59 41 11.4 &    6.0 &    2.4 &   16.0 &    4.0 &    0.7 &    2.8 &    1.0 &   B1 &   30.814  \\
\enddata

  \end{deluxetable}

\clearpage

\begin{deluxetable}{llcc}
\footnotesize
\tabletypesize{\scriptsize} 
\tablecaption{Binaries: Velocities
 \label{tbl-7}}   
\tablewidth{0pt}
\tablehead{
\colhead{Star} & 
\colhead{Ref} & \colhead{ Binary } & \colhead{ P$_{orb}$ } \\
\colhead{ } & 
\colhead{} & \colhead{ Status } & \colhead{ $^d$ } \\
  }
\startdata 

HD93129A & 4 & n & \\
HD93129B & 4 & n & \\
HD93250 & 1 & n & \\
HD93205 & 1,3 & y & 6.1 \\
HD93160 & 1   & n & \\
HD93162 & & & \\ 
HDE303308 & 1 & y: & \\ 
HD93204 & 1 & n & \\ 
-59$^0$2600 & 1 & n & \\ 
-59$^0$2603 & 1 & y & 2.2 \\ 
HDE303311 & 1 & n & \\ 
-59$^0$2635 & 2 & y & 2.3 \\ 
-59$^0$2641 & 1 & y: & 4.0 \\ 
-59$^0$2636 & 5 & y & 3.6, 5.0 \\ 
-59$^0$2628 & 1 & y: & 5.8 \\ 
HD93343 & 1 & y: & \\ 
Tr16-9 & 1 & n & \\ 
Tr16-10 & 1 & y: & \\ 
Tr16-23 & 1 & n & \\ 
Tr16-115 & 1 & n & \\ 
Tr16-3 & 1 & n & \\ 
Tr16-244 & & & \\ 
Tr16-22 & & & \\ 
Tr16-11 & & & \\ 

\enddata

References

1 Levato et al. 1991a

2 Colombo et al. 2001                       

3 Morrell  et al. 2001

4 Penny et al. 1993

5 Albacete Colombo et al. 2002
             
6 Rauw et al. 2001
                                     
  \end{deluxetable}

\clearpage

\figcaption[car.id.ps]{ The ACIS-I image with $\eta$ Car in the center. The
image has been smoothed with a 1" gaussian. Trumpler 16 is in the center of
the image; Trumpler 14 is in the upper right. A few of the sources are
identified. The 16 x 16 arcmin ACIS-I field is shown. N is up; E is on the
left. \label{fig1}}

\figcaption[srcall.ps]{ Sources found in the $\eta$ Car image.
Sources identified with O or B stars are marked with circles;
other sources are diamonds.  All the symbols are the same size, 
although the source point spread function is larger off axis. 
 \label{fig2}}

\figcaption[imob.ps]{The $\eta$ Car image with the positions of
known  O and early B stars overlaid (circles). 
 \label{fig3}}

\figcaption[obdet.ps]{O and B stars on the $\eta$ Car ACIS-I image.
Spectral type and V magnitude are taken from Massey and Johnson (1993).
Squares are detected X-ray sources; Xs are not detected in X-rays. 
Spectral types classes are sequential from 0 to 9 for O stars,
and 10 to 15 for B stars.
 \label{fig4}}

\figcaption[obcts.ps]{Hot stars detected on the $\eta$ Car ACIS-I image. 
Counts are the total counts.  The Wolf-Rayet star (log cts = 3.7) has 
been arbitrarily plotted at O7 for comparison
 \label{fig5  }}

\figcaption[viwolk.ps]{V image of the center of the ACIS-I image.  
 This image is          
 approximately 12 x 12 arcmin centered on $\eta$ Car.  The circles               
 are the positions of X-ray sources, omitting X-ray sources that have been       
 identified with OB stars.  Most, but not all, of the circles
contain a star,  indicating           
 that a cool star is the source of the X-rays.                                   
 Image credit: Nuria Huelamo (MPE), Takashi Isobe (CfA) and the 0.9              
 meter at CTIO.                
 \label{fig6  }} 

\figcaption[tr14_tr16.ps]{The V vs V-I color magnitude diagram from the field
in Fig. 6. X-ray sources associated with Tr 16 are marked with *; 
X-ray sources associated with Tr 14 are marked with circles;
other stars are dots. The dots define a main sequence presumably
of field stars in the Carina arm.  The X-ray sources form a sequence 
which is more luminous than the field star sequence for cooler stars,
but coincides with it for hotter stars.  This is what would be expected if the 
X-ray sources are  pre-main sequence stars.  V and (V--I) are
in magnitudes in all figures.   \label{fig7 }}


\figcaption[obhist.ps]{Luminosity distribution of OB stars.
Luminosity is in ergs sec$^{-1}$.
\label{fig8 }}

\figcaption[lumnob.ps]{Luminosity distribution of sources which are 
not identified as O and early B stars. Luminosity is in ergs sec$^{-1}$.
\label{fig9 }} 


\figcaption[fxfb.ps]{The log of the ratio of X-ray luminosity to 
bolometric luminosity as a function of spectral type for O and early B stars.
O and B stars are squares; the Wolf-Rayet star is an asterisk, plotted
at an arbitrary spectral type.
\label{fig 10}}

\figcaption[coollx.ps]{The log of the X-ray luminosity as a function of 
V--I.   Luminosity is in ergs sec$^{-1}$.  The vertical lines represent
(from left to right) the colors of  A0 V, F0 V, and K7 V stars with the 
reddening of the clusters. 
\label{figxxx_a }} 

\figcaption[coolrat.ps]{The log of the ratio of the  X-ray luminosity
to the bolometric luminosity as a function of 
V--I.   Luminosity is in ergs sec$^{-1}$.  The vertical lines represent
(from left to right) the colors of  A0 V, F0 V, and K7 V stars with the 
reddening of the clusters. 
\label{figxxx_b }} 

\figcaption[ecar_diffuse_regions.ps]{Smoothed image of the field
to bring out the diffuse emission across the observation.  
The adaptive smoothing routine {\tt CSMOOTH} was used.  Circles
indicate regions in which diffuse image is discussed in the text.
\label{fig11 }} 

\figcaption[ndet.ps]{Hot stars {\it not} detected on the $\eta$ Car 
ACIS-I image.  Circles are 30" in diameter.      
 \label{fig12}}

\figcaption[ulcts.ps]{Counts at the positions of undetected hot stars.
The squares are the counts within 2.5 arcsec circles as a function of 
radial distance from the telescope axis (in arc min).  Filled triangles
show our detection criterion (see text) corresponding to each 
undetected source.
 \label{fig13  }}

\figcaption[ntcts.ps]{Net counts at the positions of undetected hot stars.
The squares are the net counts within 2.5 arcsec circles as a function of 
radial distance from the telescope axis (in arc min).  (One point with very
large net counts is omitted.) The approximate division between 
 stars in Tr 16 (near the field center) and 
stars in Tr 14 is indicated.
 \label{fig14  }}

\figcaption[ullum.ps]{Upper limit to the 
log of the luminosity in ergs cm$^{-1}$ for 
undetected hot stars as a function of spectral type. 
x's are stars more than 7' from the aimpoint
(Tr 14); squares are stars closer than 7' (Tr 16).
 \label{fig15  }}

\figcaption[hard.ps]{The hardness ratio ``color-color" diagram.
HRH is the hardness ratio from the hard and medium energy 
bands; HRS is the hardness ratio from the medium and soft bands.
See text for discussion.
Squares are OB stars; x's are other sources,  presumably 
pre-main sequence stars.  The solid line shows hardness ratios 
created using PIMMS for Raymond-Smith plasma for a 
variety of temperatures for solar abundance; the dashed line is for 
0.2 solar abundance.  The dot-dash line is from thermal bremsstrahlung, 
also for a range of temperatures.   The dotted 
lines shows the effect of varying the extinction (Raymond-Smith, 
solar abundance) from 
N$_H$ = 3 x 10$^{21}$ to 5  x 10$^{21}$, corresponding approximately to 
the range of E(B-V) in the field from 0.5 mag (standard) to 0.9 mag 
(exceptional).  The dotted line near HRS -0.2 has been raised by 
0.05 in HRH, so that it is not completely hidden by the thermal
locus.  
 \label{fig 16 }}

\figcaption[hardap.ps]{The soft hardness ratio, HRS, as a function of 
spectral type for the hot stars.  Squares are O stars; the Wolf-Rayet
star has been arbitrarily plotted as an * at O7.  Soft colors are at 
the bottom.  
 \label{fig17 }} 

\clearpage


\plotone{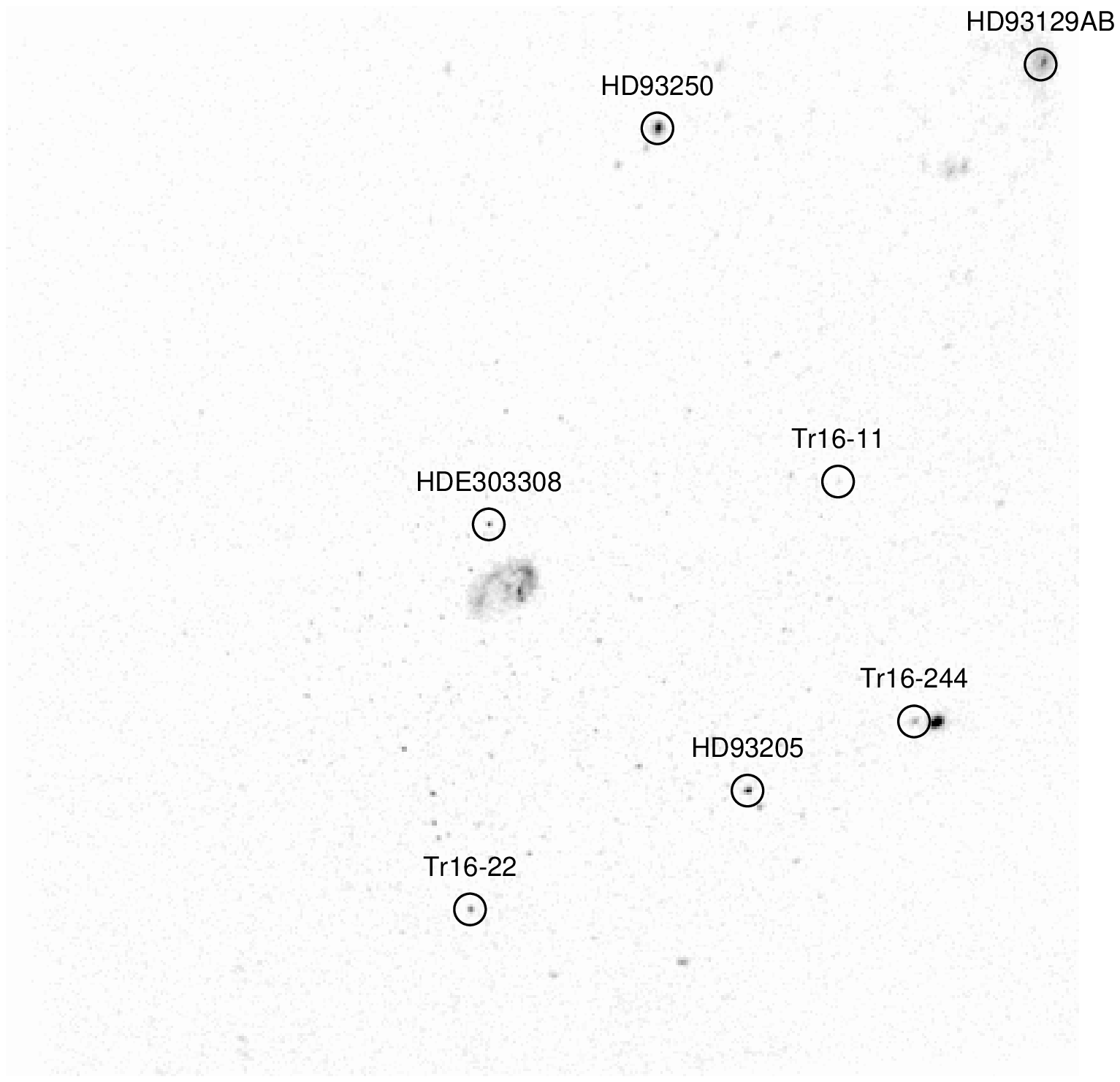}

\plotone{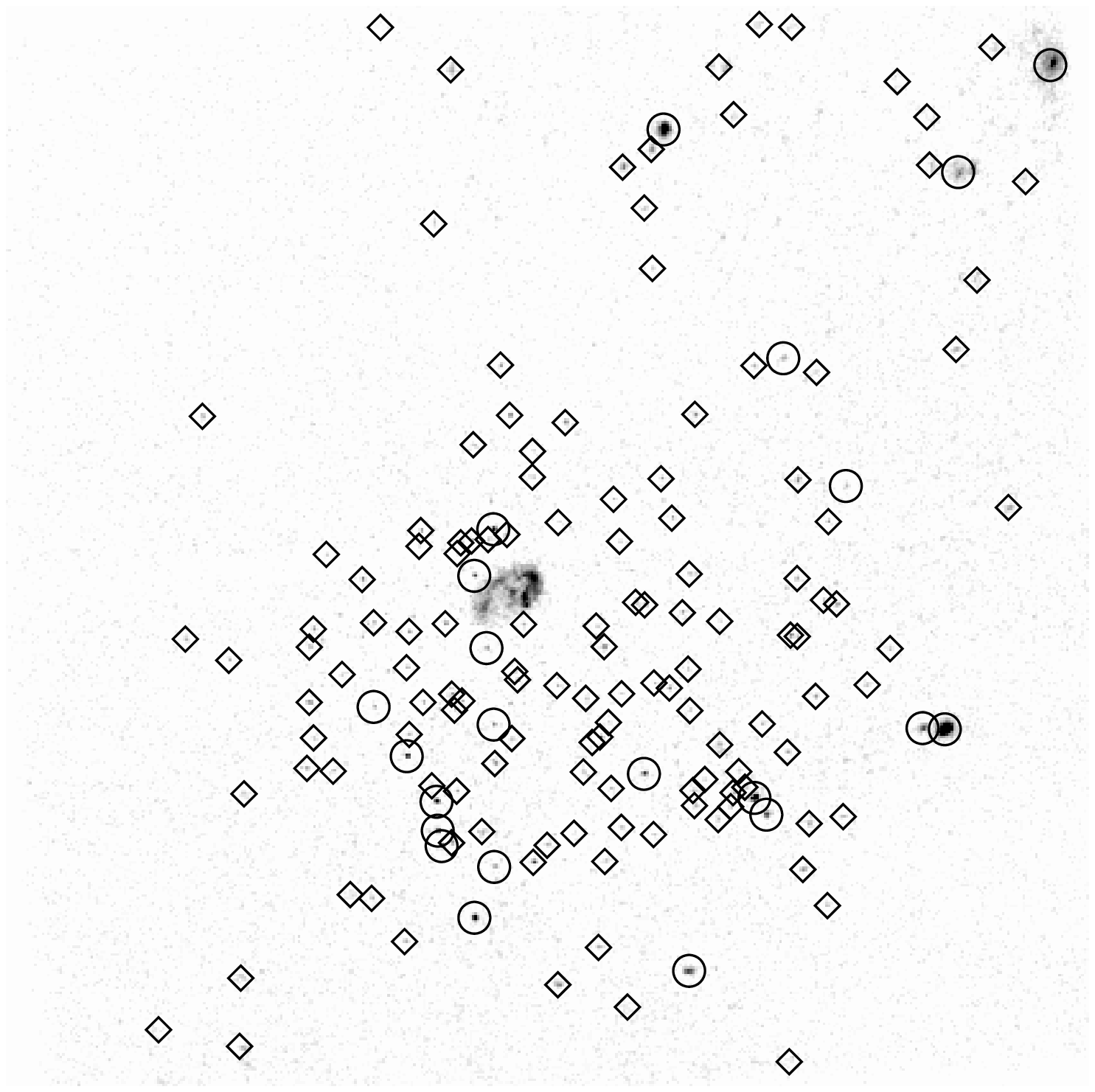}
                      
\plotone{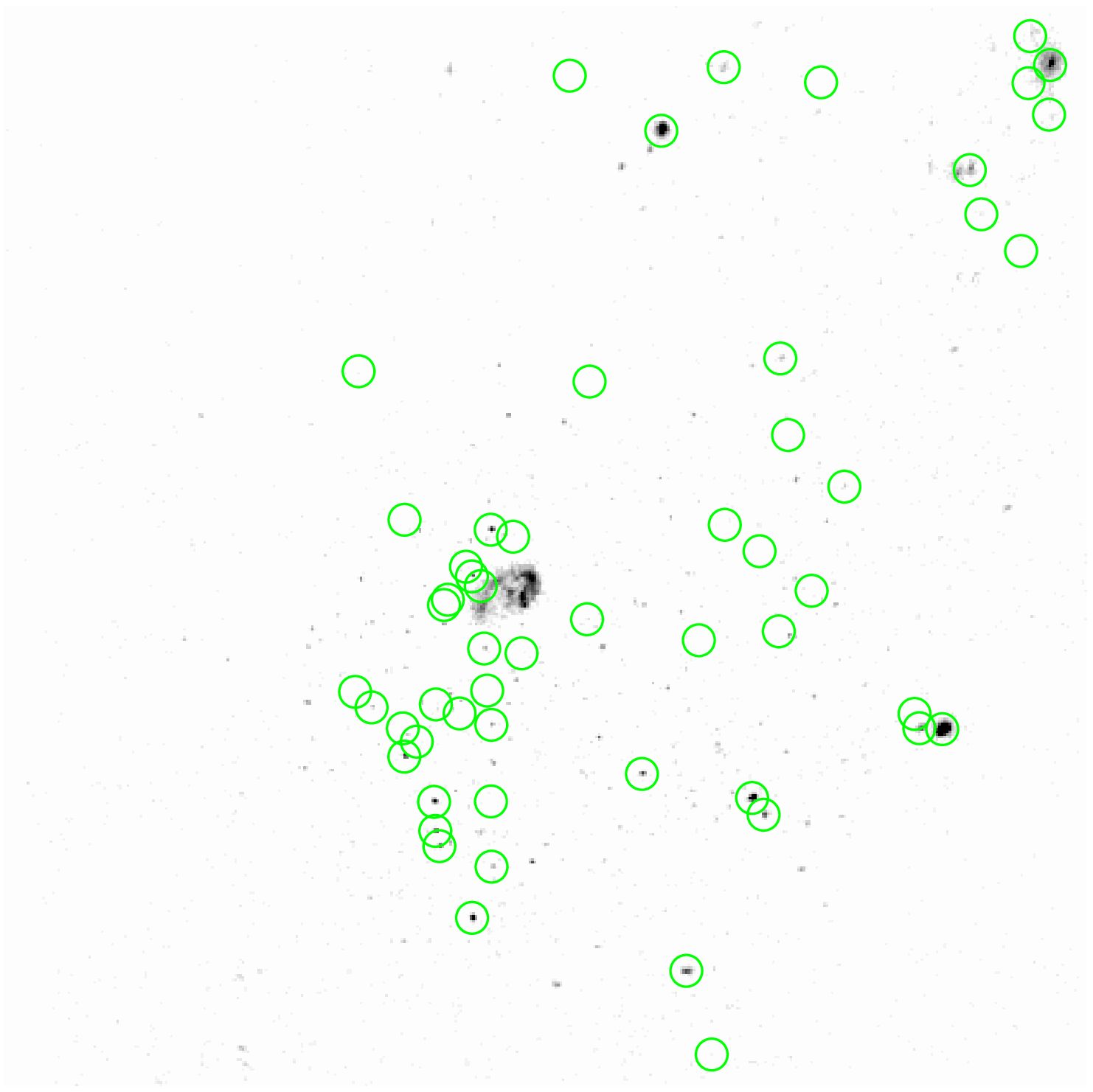}

\plotone{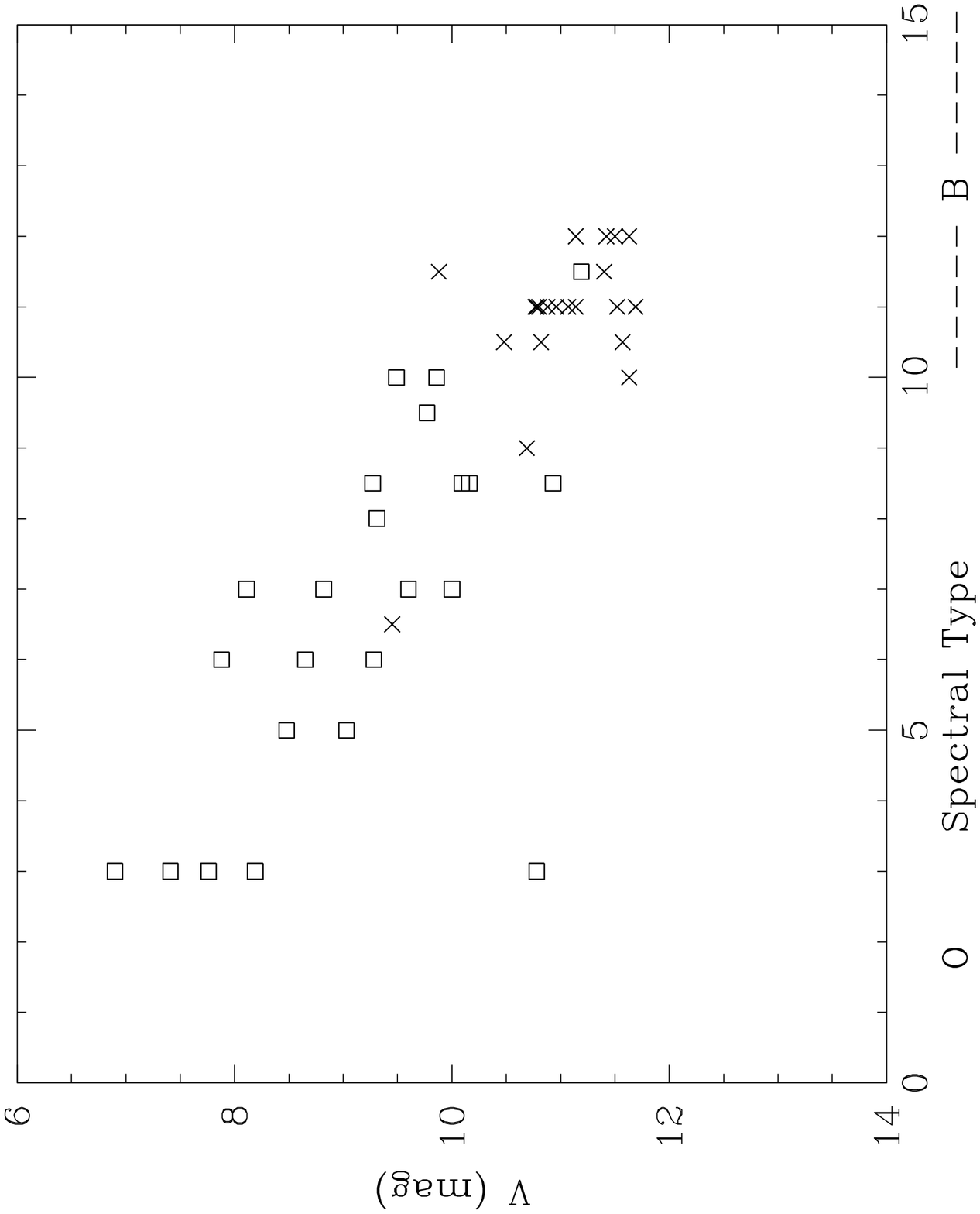}
          
\plotone{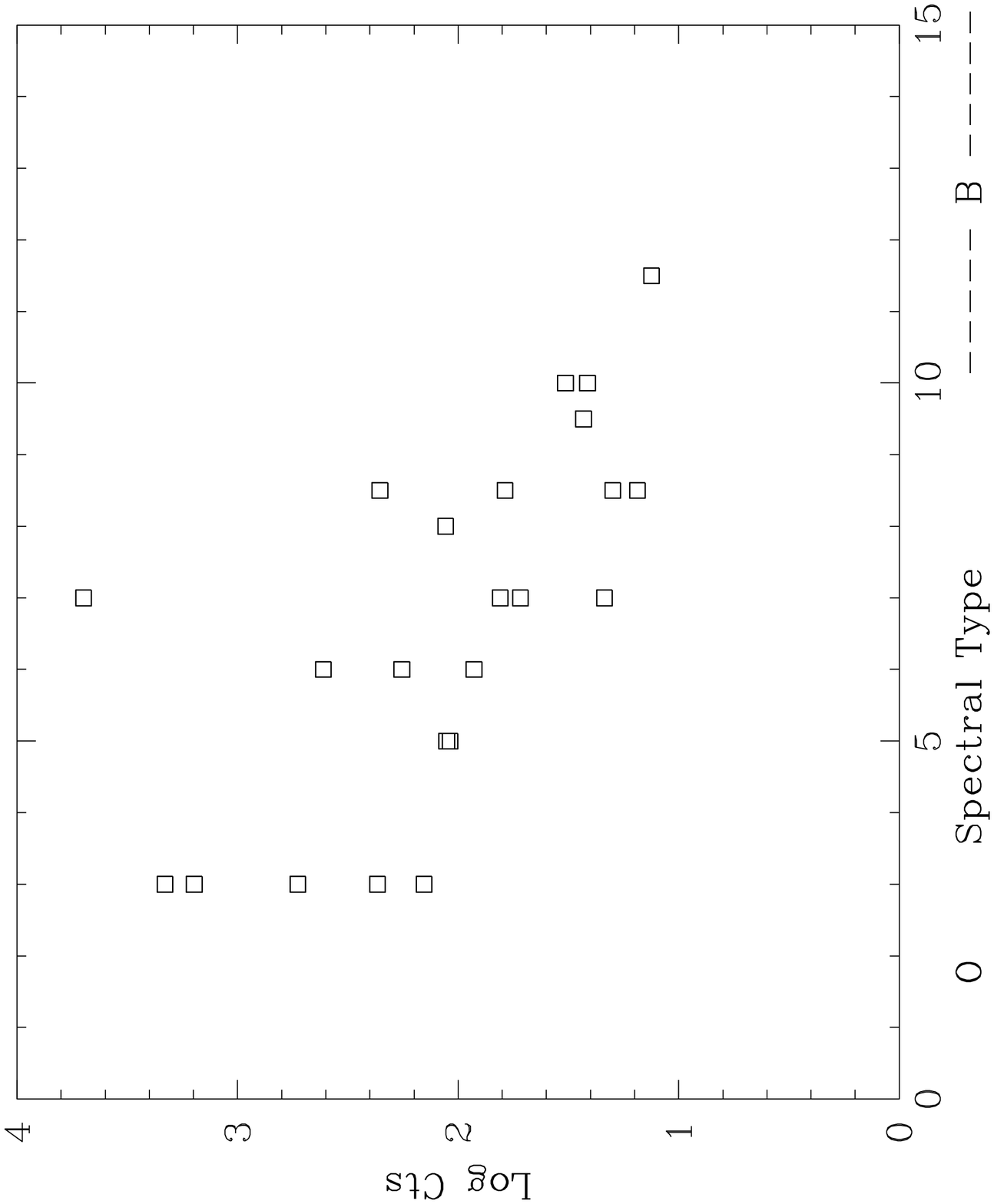}             
             
\plotone{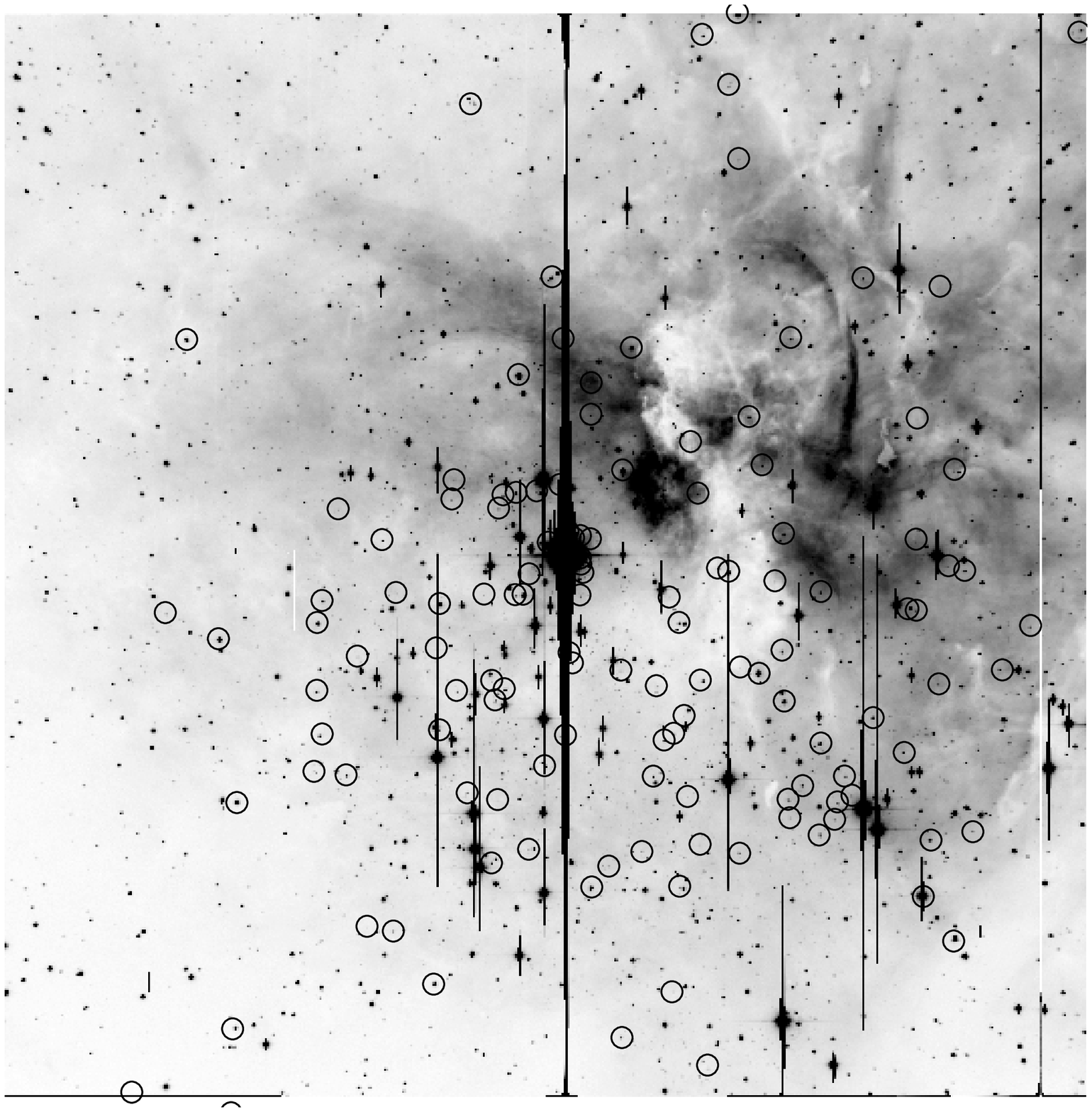}


\plotone{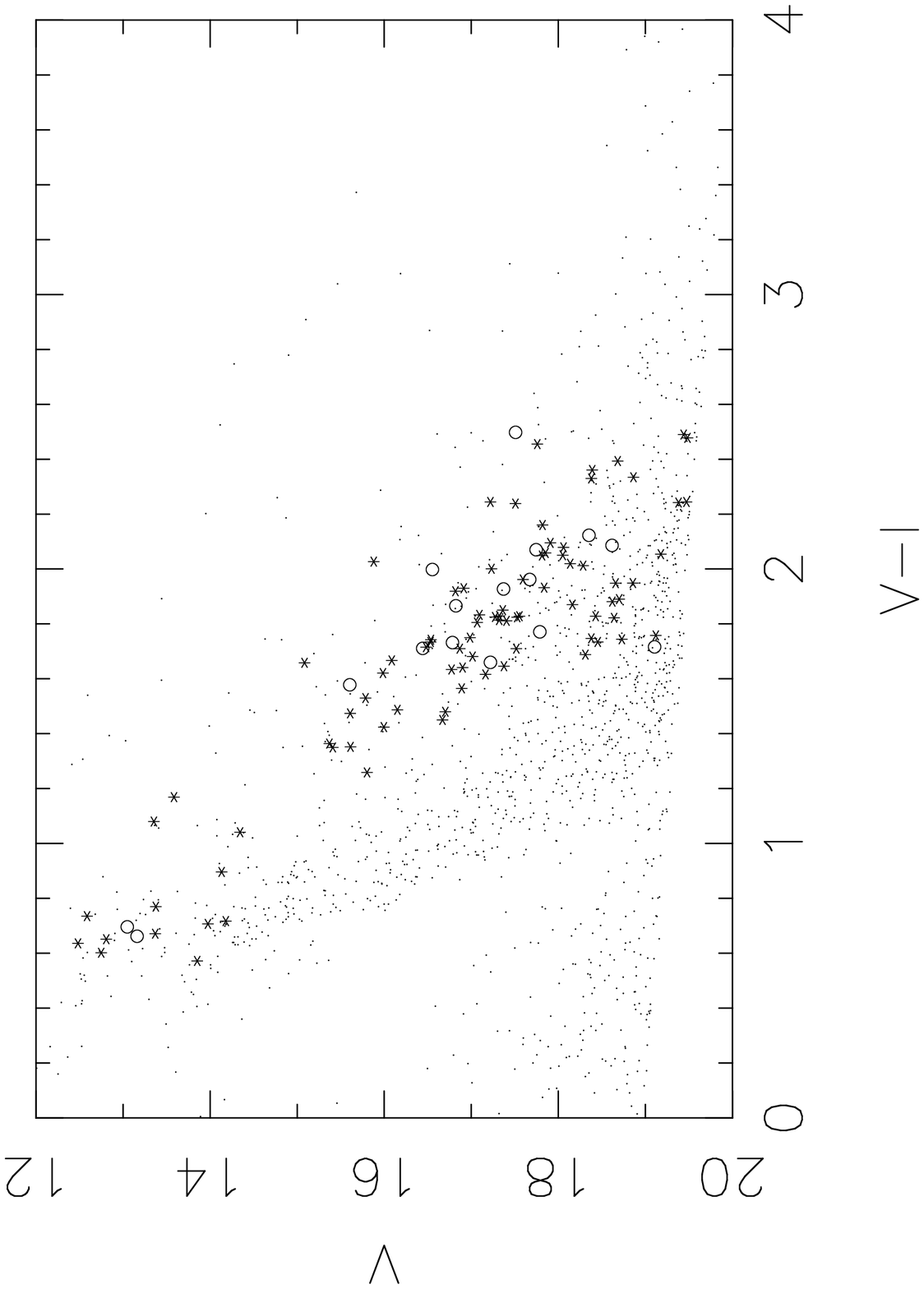}

\plotone{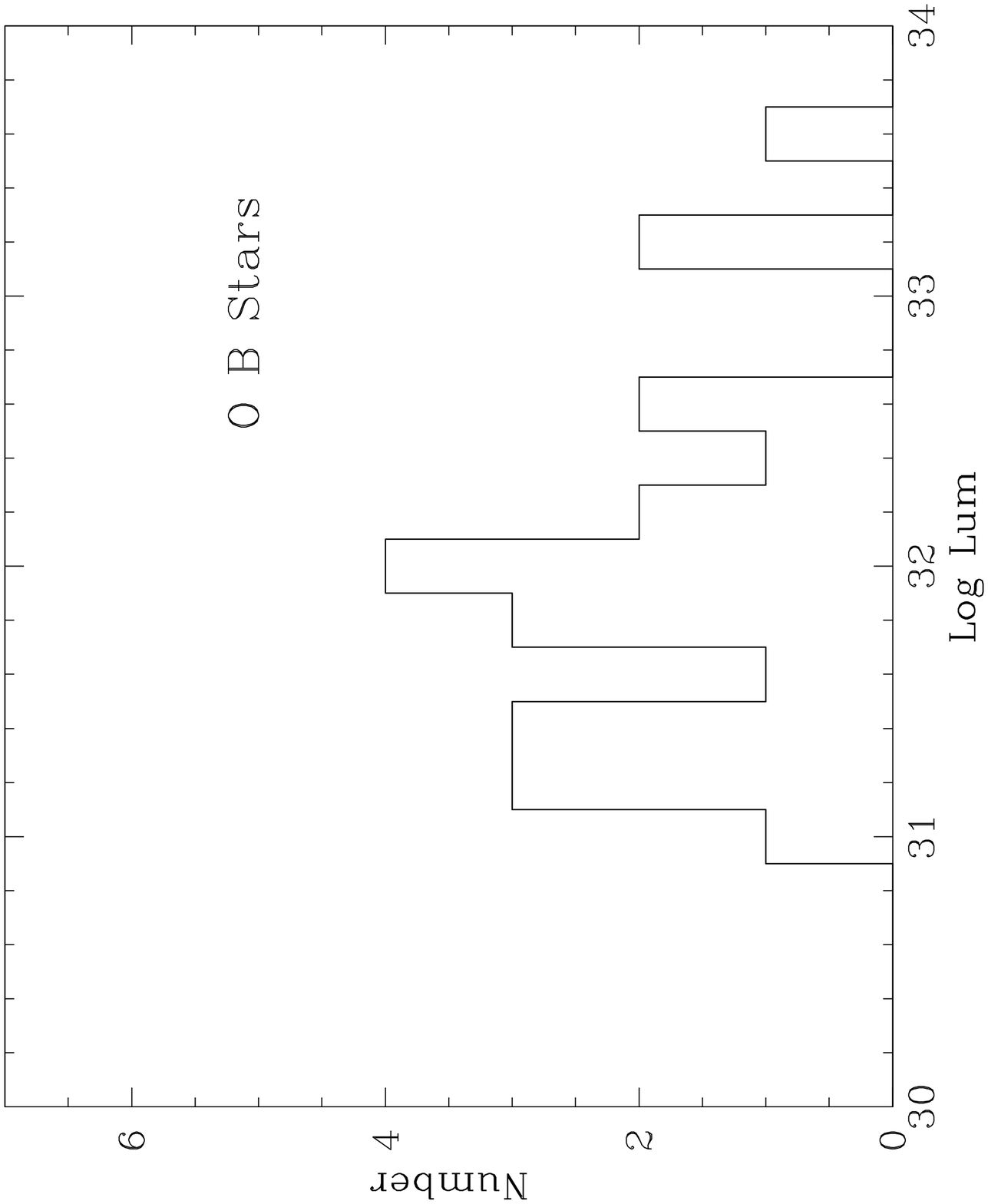} 


\plotone{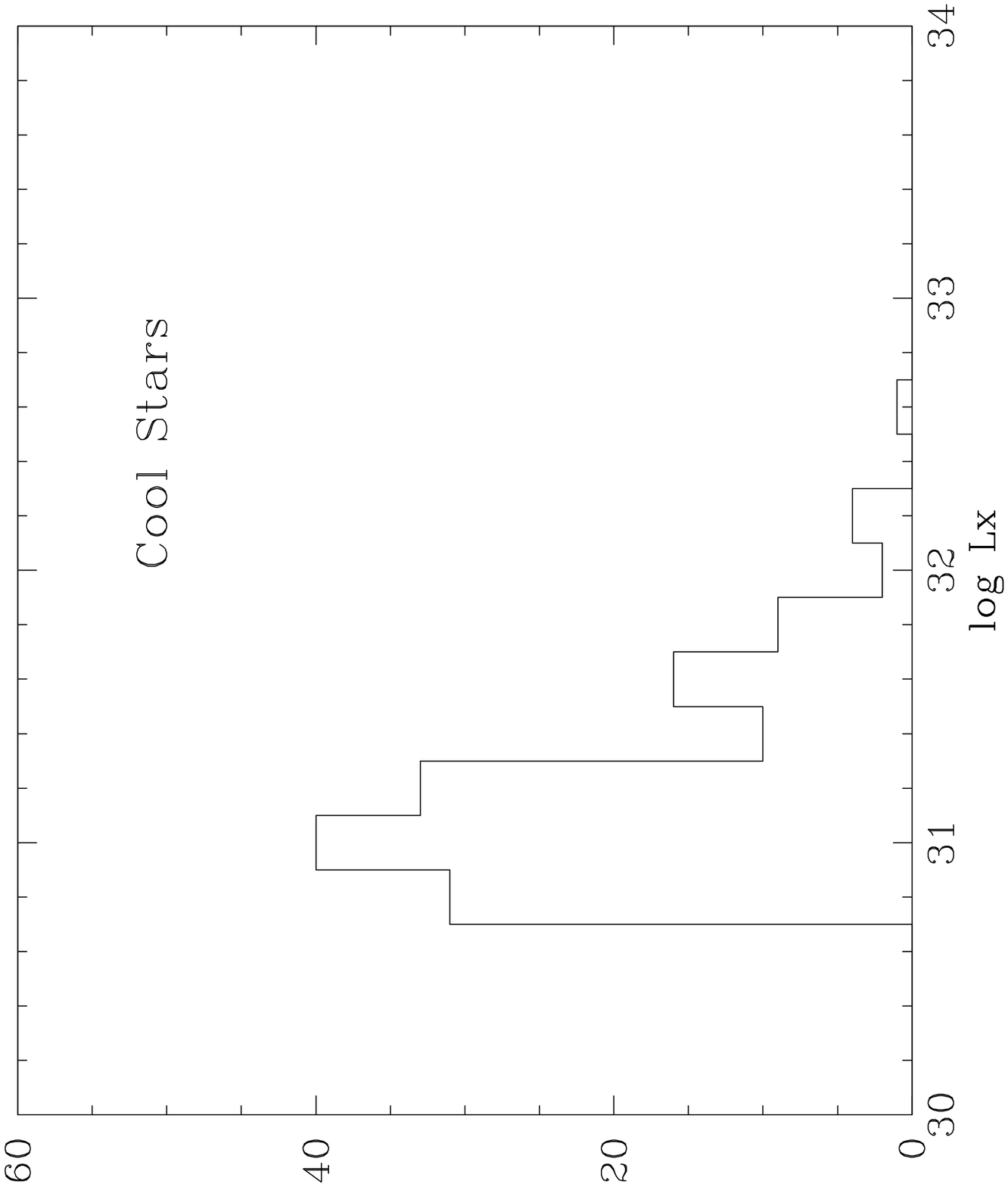}


\plotone{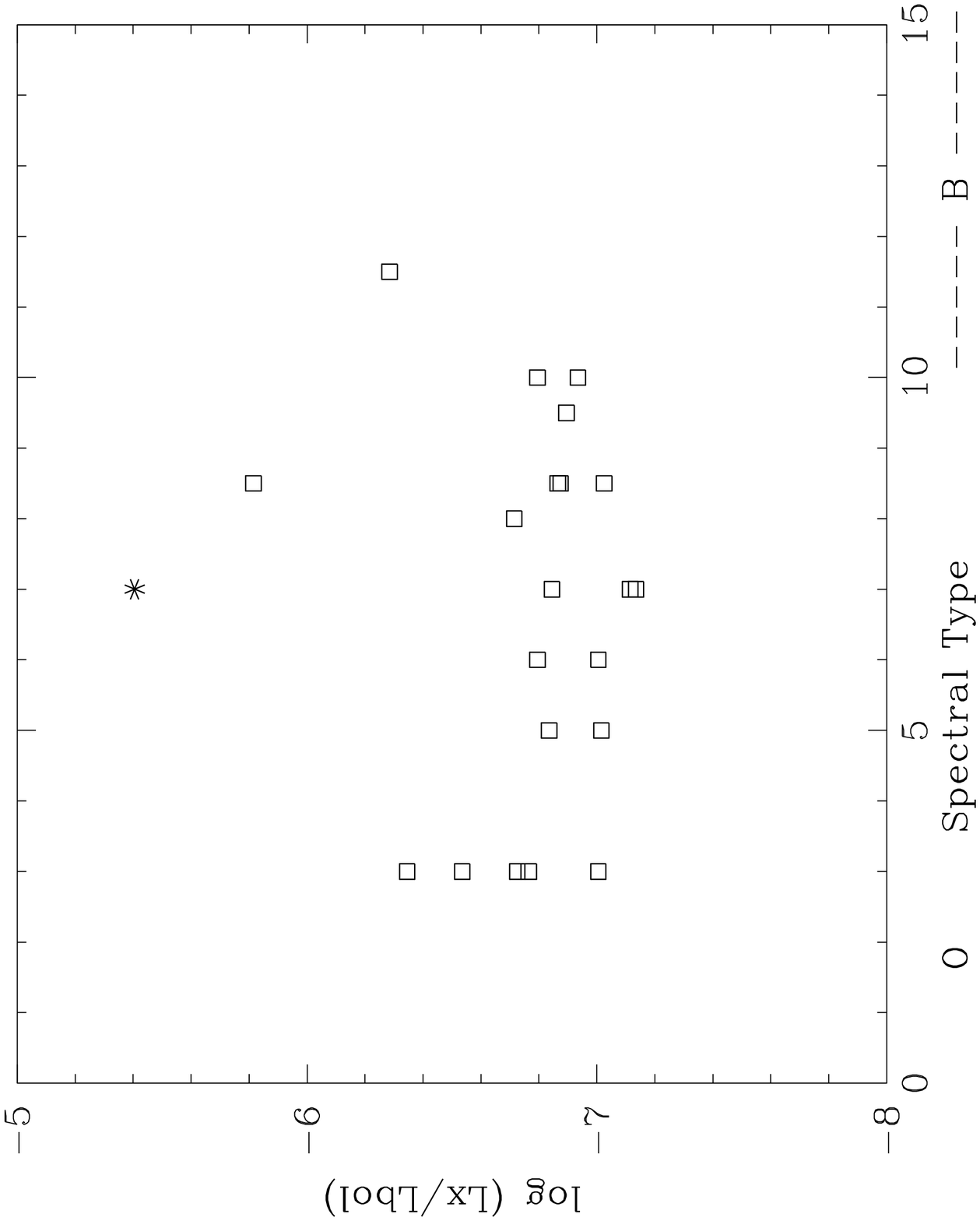}

\plotone{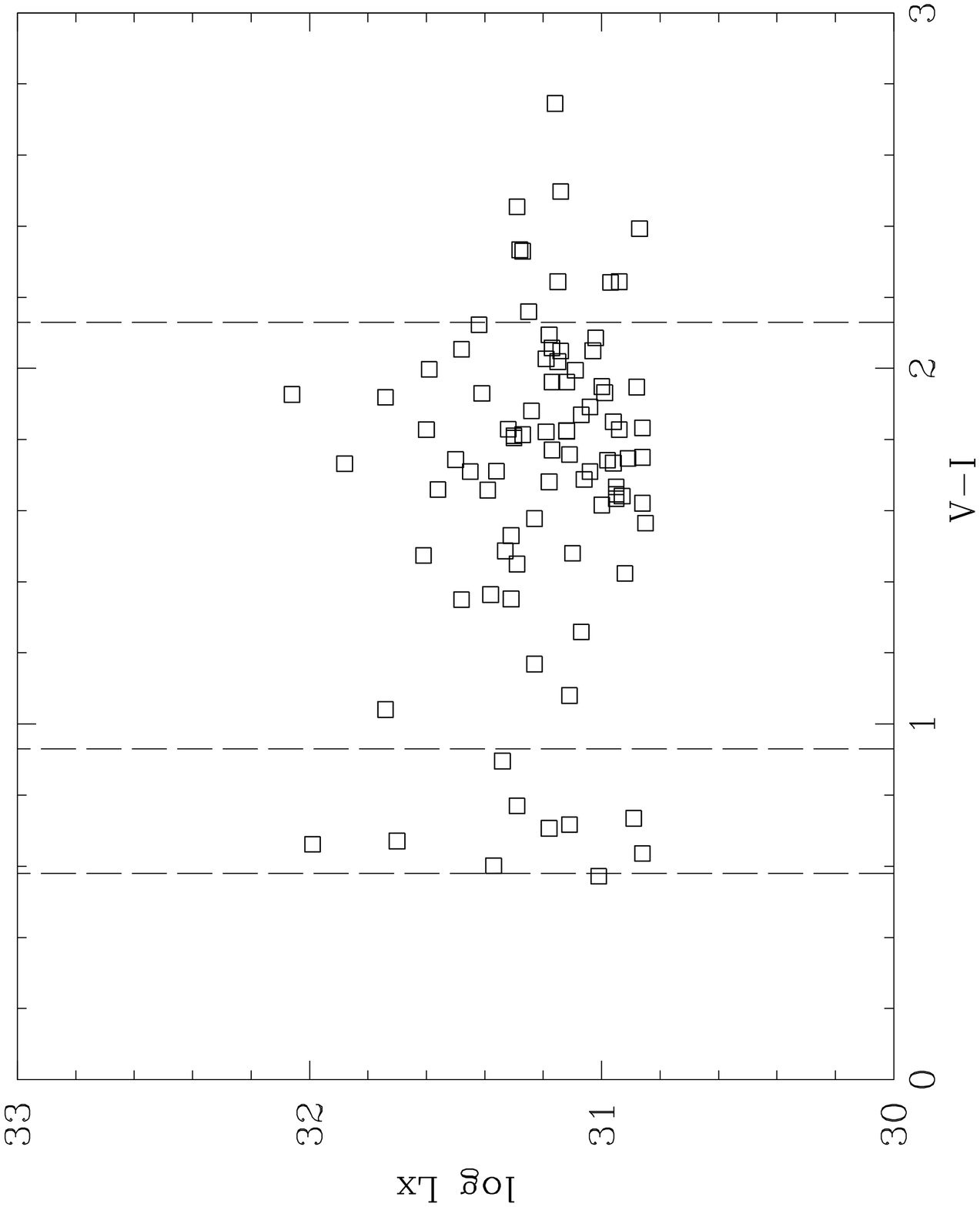}

\plotone{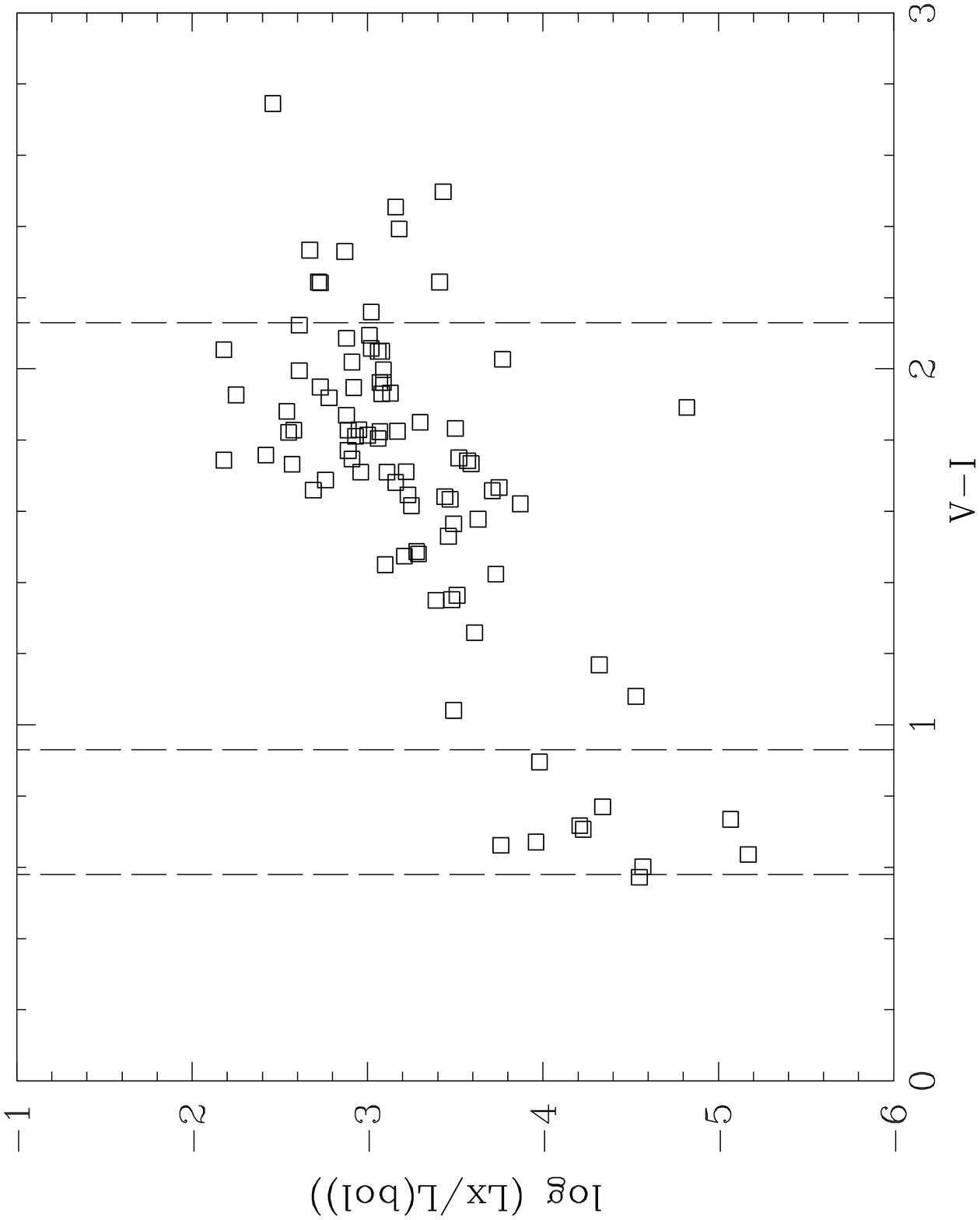}

\plotone{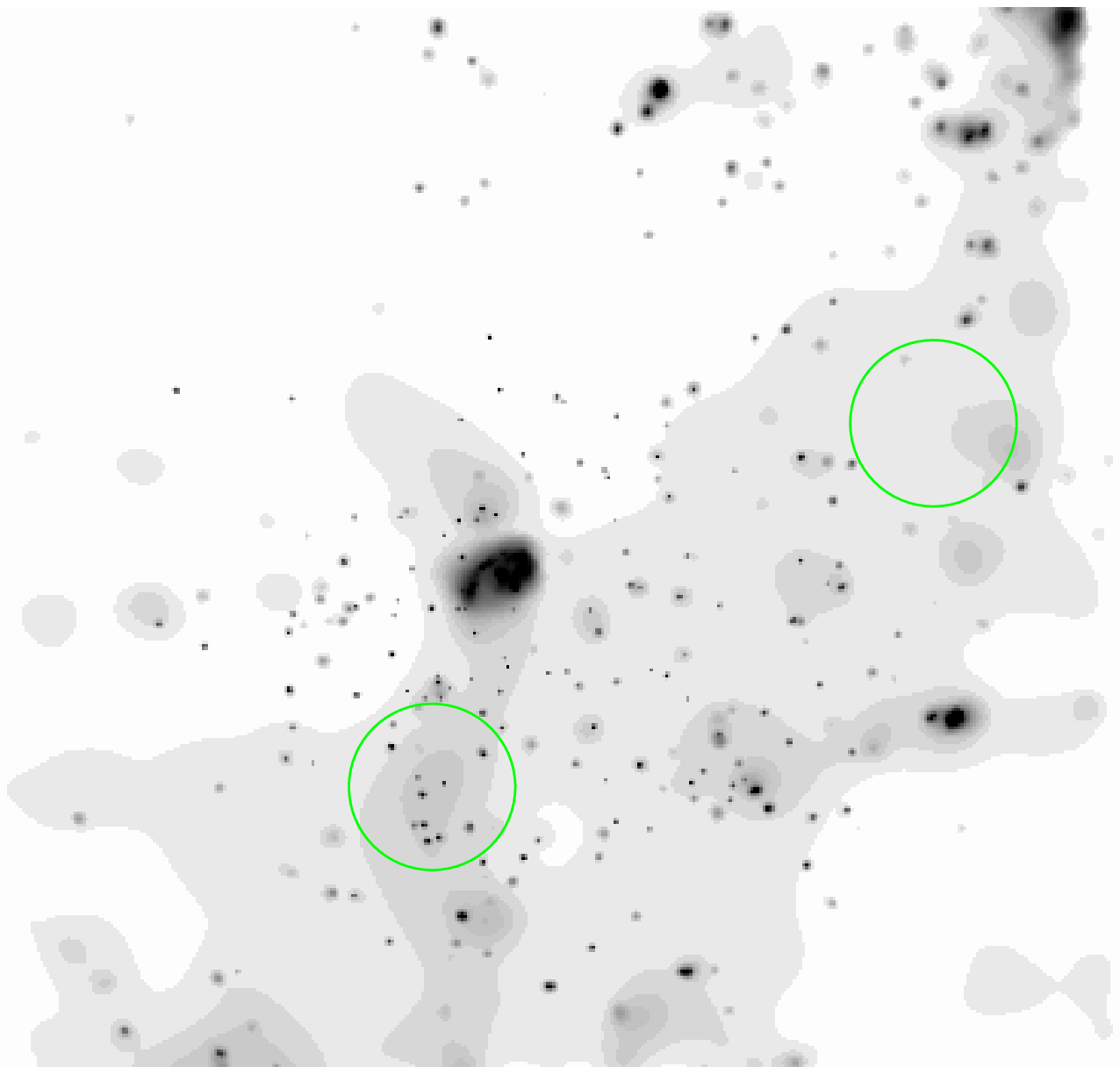}

\plotone{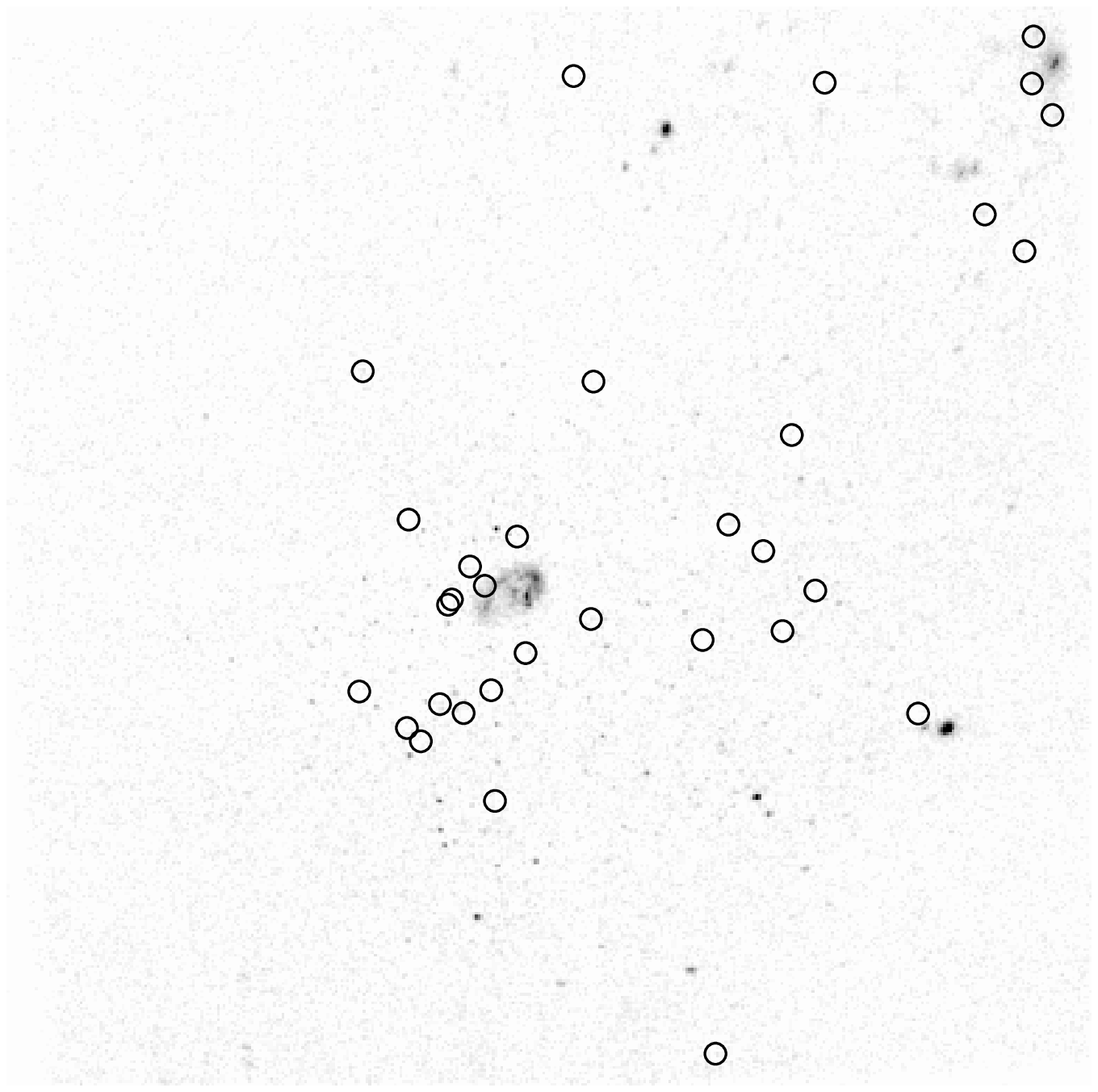}             
             
\plotone{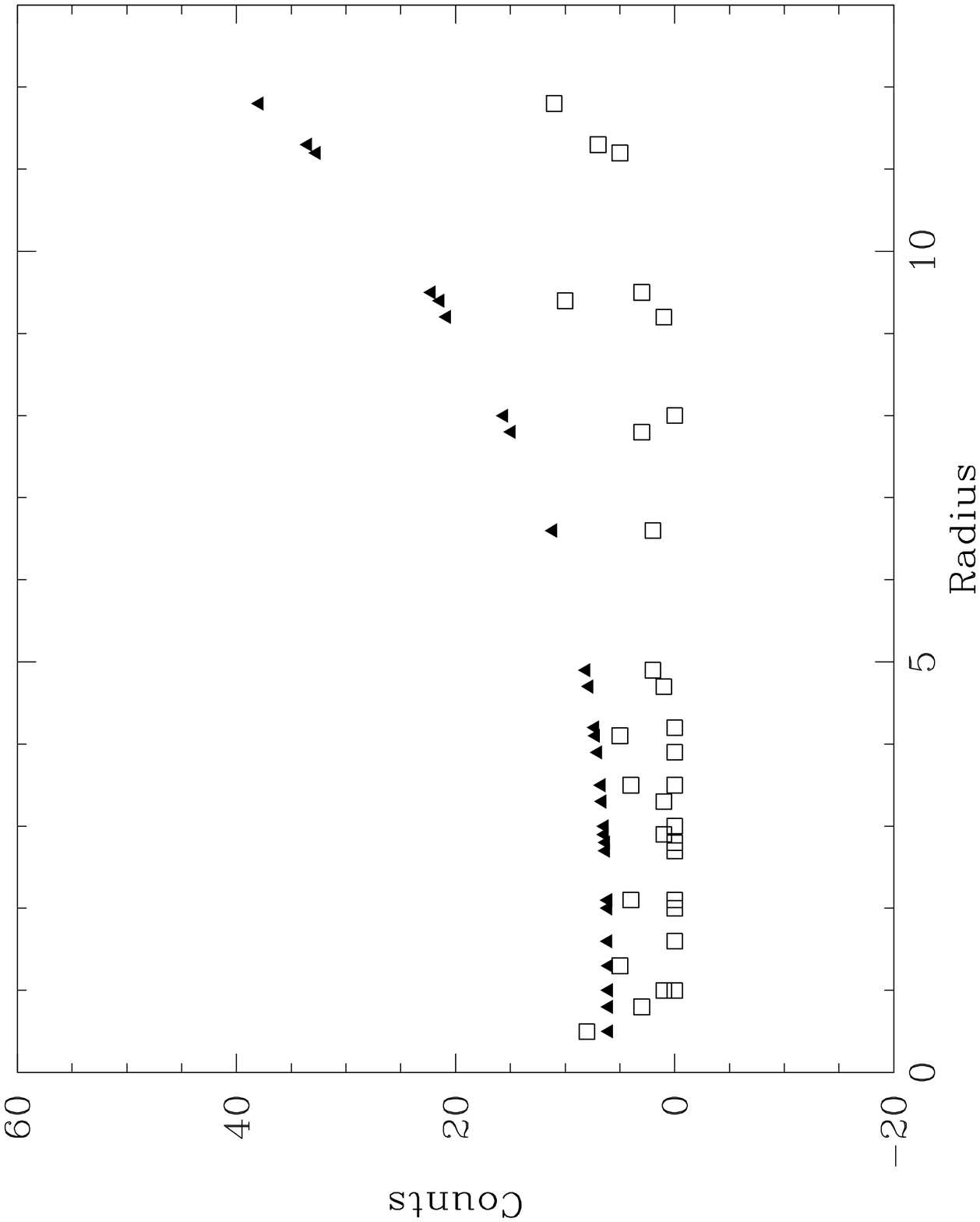}
             
\plotone{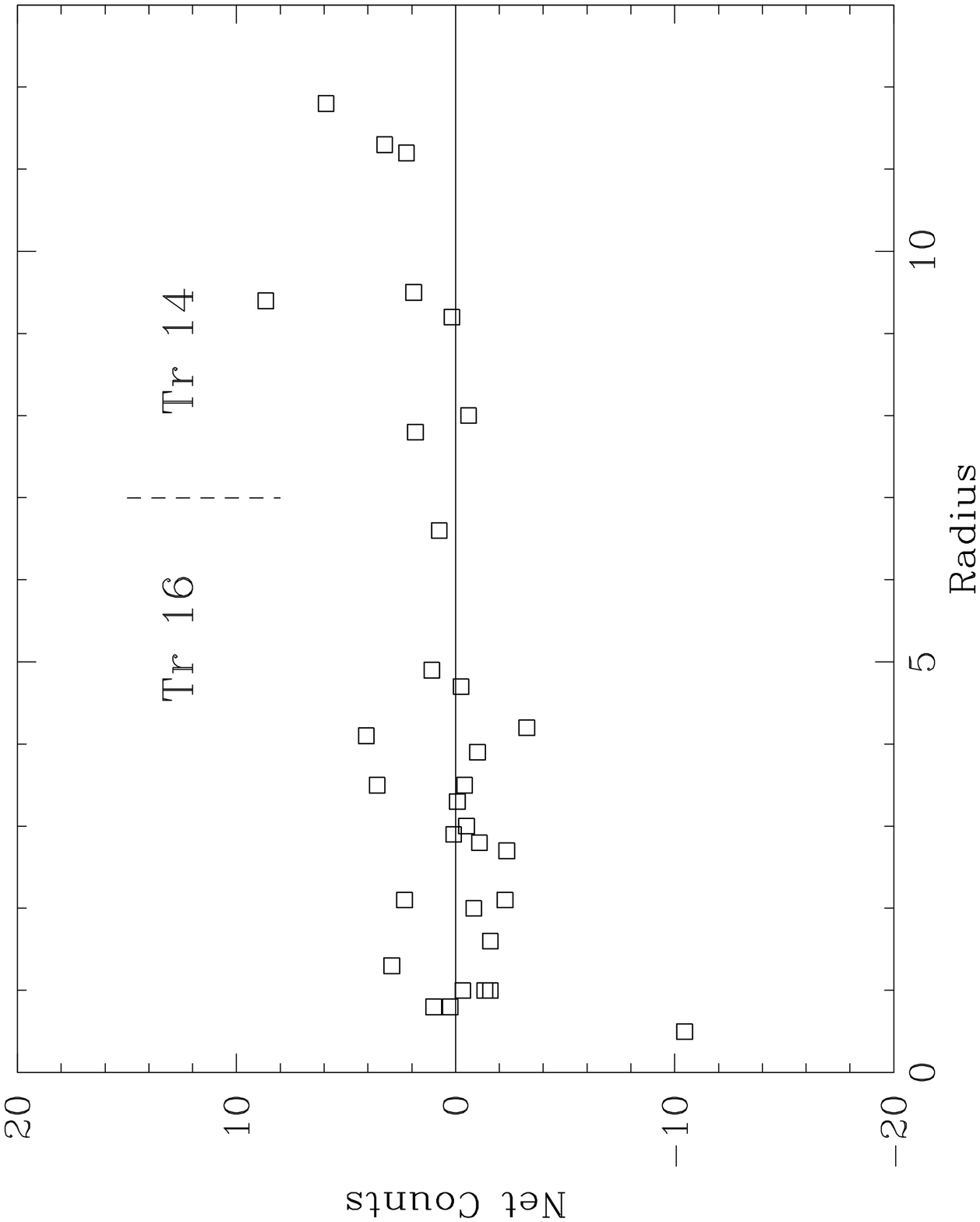}
 
\plotone{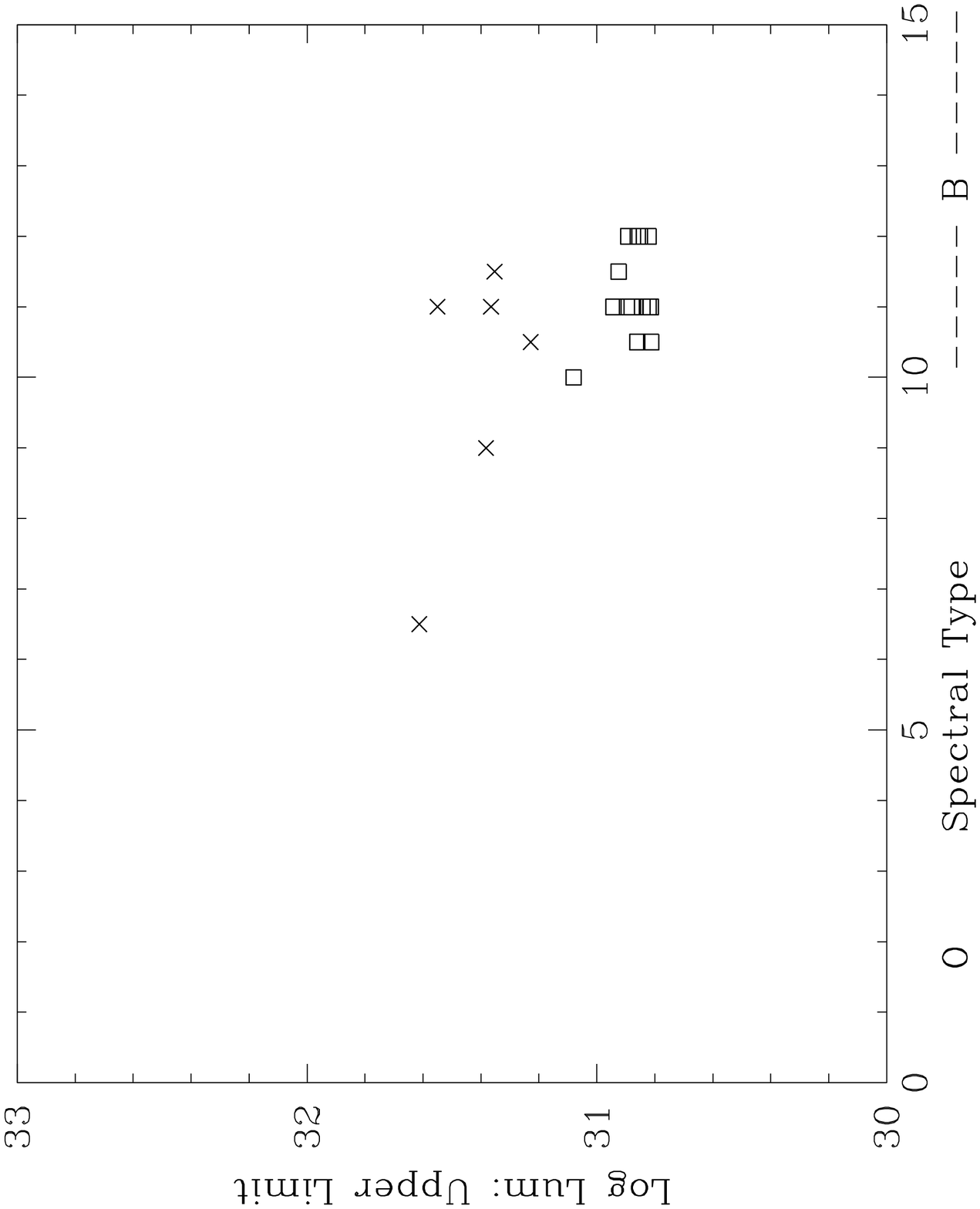}

\plotone{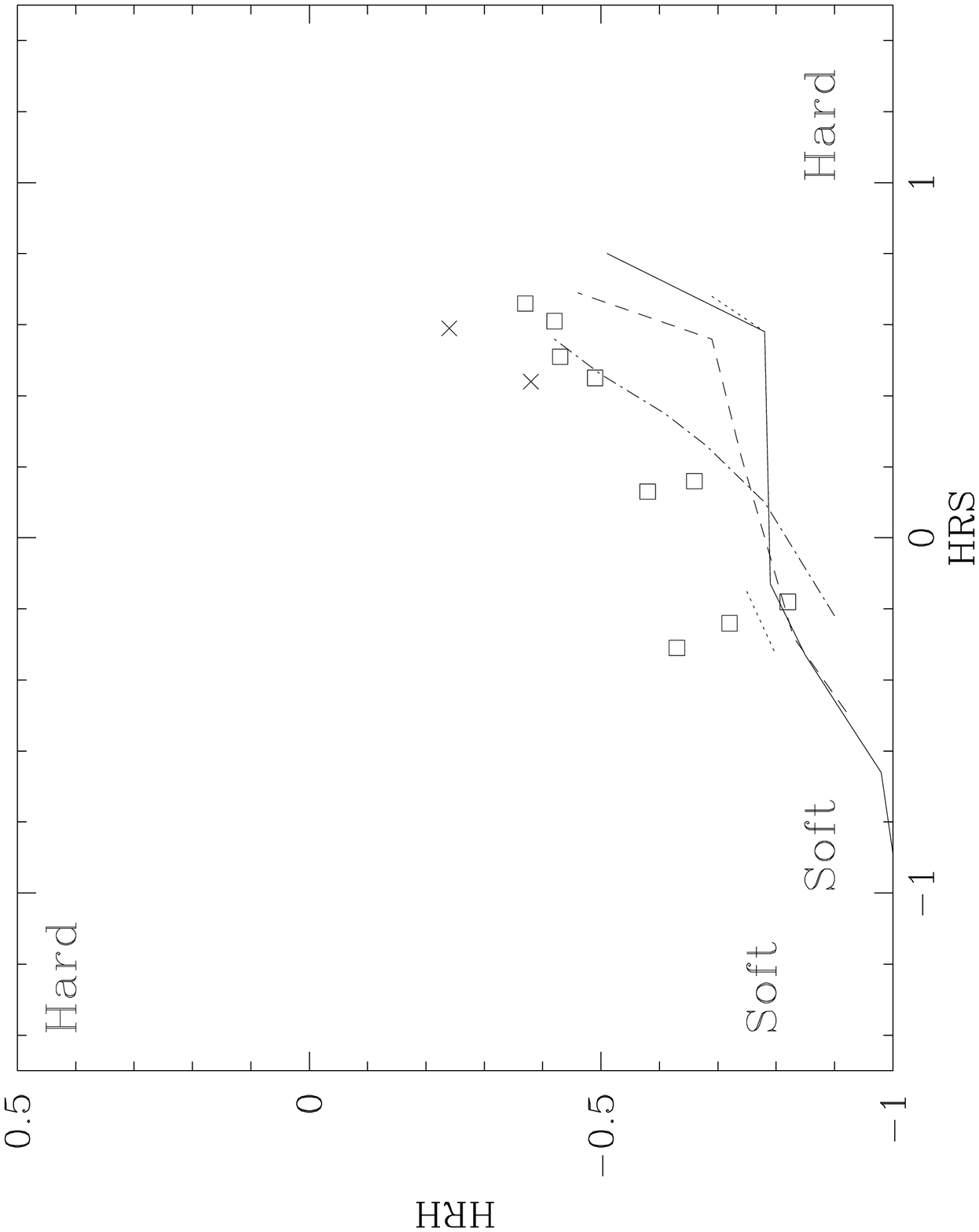}

\plotone{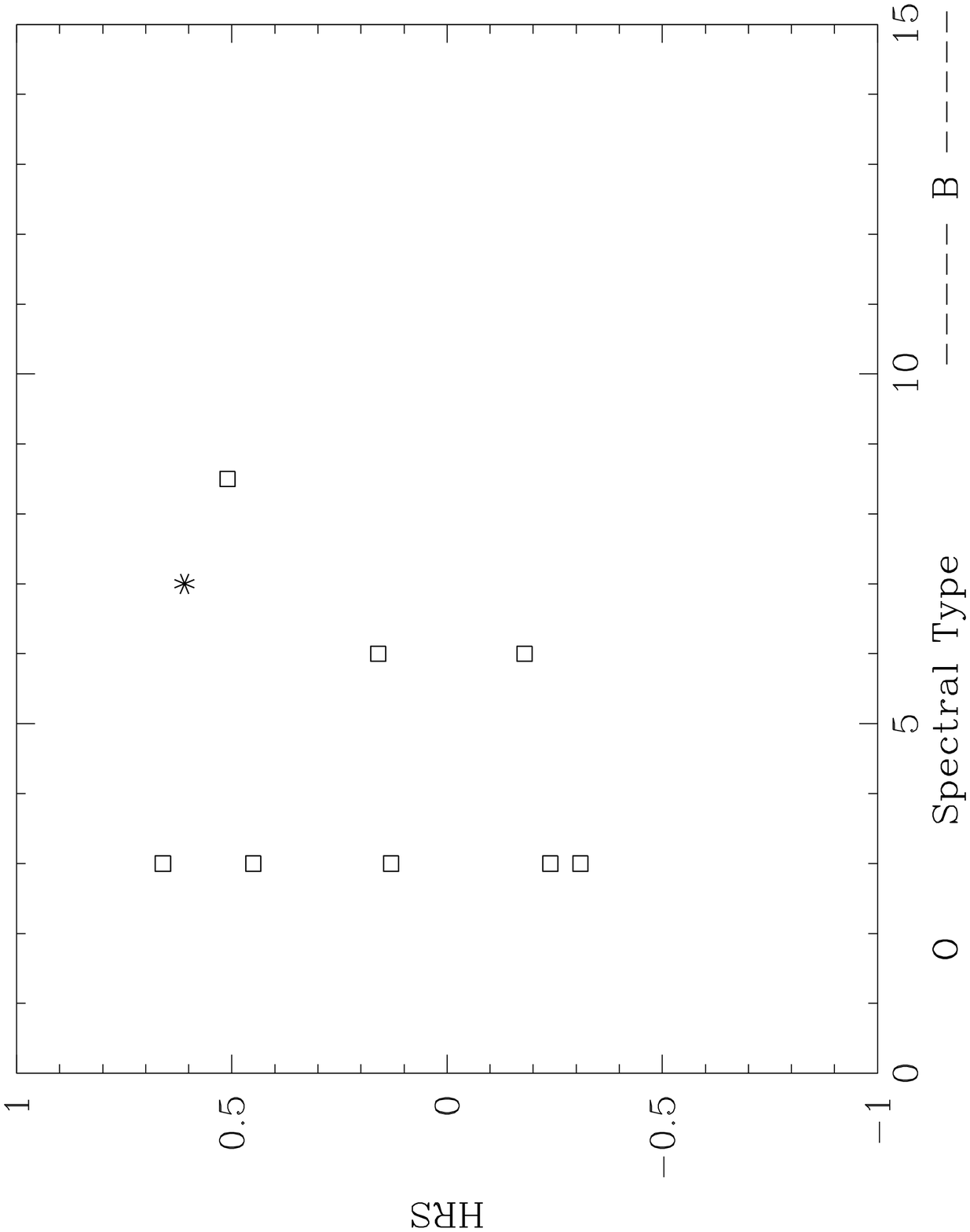}

\end{document}